\documentclass[prb,twocolumn,showpacs,showkeys]{revtex4-1}

\usepackage{amsmath}    
\usepackage{graphicx}
\usepackage{verbatim}   
\usepackage{color}      
\usepackage{subfigure}  
\usepackage{hyperref}   
\usepackage{blindtext}

\begin{document}

\title{Superfluid drag in the two-component Bose-Hubbard model}	

\author{Karl Sellin}
\affiliation{Department of Physics, The Royal Institute of
  Technology, SE-10691 Stockholm, Sweden}  
 \author{Egor Babaev}
\affiliation{Department of Physics, The Royal Institute of
  Technology, SE-10691 Stockholm, Sweden}  
\date{\today}

\begin{abstract}
In multicomponent superfluids and superconductors, co- and counter-flows of components 
 have in general different properties.
It was discussed in 1975  by Andreev and Bashkin, in the context of
 He$^3$/He$^4$ superfluid mixtures, that inter-particle   interactions produce a dissipationless
  \textit{drag}. The drag can be understood as a superflow of one component induced by phase gradients of the other component. Importantly the drag can be both positive (entrainment) and negative (counter-flow).
  The effect is known to be 
of crucial importance  for many properties of diverse physical systems ranging from  the
dynamics of neutron stars, rotational responses of Bose mixtures of ultra-cold atoms
to magnetic responses of multicomponent superconductors.
Although there exists a substantial literature that 
includes the drag interaction phenomenologically, much fewer regimes are covered by quantitative studies
of the microscopic origin of the drag and its dependence on microscopic parameters.
Here we study the  microscopic origin and strength of the drag interaction in a 
quantum system of two-component 
bosons on a lattice with short-range interaction.
 By performing quantum Monte-Carlo simulations of a two-component Bose-Hubbard model we
 obtain dependencies of   the drag strength  on the 
boson-boson interactions and properties of the optical lattice. 
Of particular interest are the strongly-correlated
regimes where the ratio of co-flow and counter-flow superfluid stiffnesses can diverge,
corresponding to the case of saturated drag.

\end{abstract}

\maketitle

\section{Introduction}
Superfluids are in general multi-component systems and as such are
 characterized by a matrix of superfluid stiffnesses.
  The matrix describes
 superflows of individual components 
as well as their co-flow and relative motion.
Since the particles comprising superfluids in general have interspecies interaction,
their superflows will be interacting as well.
As a result the co-flow of components is
different from the counterflow. The effect was 
 first discussed by Andreev and Bashkin in the context of 
 superfluid mixtures of He$^3$ and He$^4$ isotopes: namely that there
 will be a dissipationless  drag between superflows of two components \cite{andreev1976three}. 
 The intercomponent entrainment (dissipationless drag) is often referred to as the
 Andreev-Bashkin effect.
  Later it was realized that  the effect  has many  important consequences in a wide variety of systems.
  In nuclear Fermi liquids there is an entrainment effect between the neutronic superfluid and
 the protonic superconductor, which  is a crucial part of the current models of observed dynamics of neutron stars  \cite{alpar1984rapid,SJOBERG1976511,PhysRevC.54.2745,BALDO1992349,Chamel,link2003constraining,babaev2004andreev,jones2006type,PhysRevB.78.024510,PhysRevLett.103.231101}. 
The entrainment effect has also been argued to be rather generically
 present and important for the physical properties of 
 triplet superconducting and superfluid  states \cite{RevModPhys.47.331}.
 In particular it was discussed to lead to a stabilization of 
half-quantum vortices and skyrmions in superconducting  systems
 \cite{PhysRevLett.99.197002,chung2009fractional,garaud2014skyrmions}.

Recently it was realized that the drag effect is especially important
in
strongly correlated  superfluid mixtures in optical lattices.
Even a weak drag  interaction substantially affects the vortex states in such systems \cite{dahl2008preemptive,PhysRevB.78.144510,PhysRevLett.101.255301}.
A sufficiently strong interaction leads to the appearance of new phases, namely, it is possible to have phase transitions to states where only co-flows exist (paired superfluids) or only counter-flows exists (super-counterfluids)  \cite{kaurov2005drag,kuklov2004commensurate,kuklov2004superfluid,dahl2008preemptive,PhysRevB.78.144510,PhysRevLett.101.255301,svistunov2015superfluid,PhysRevLett.90.100401,kuklov2006deconfined}.
These phase transitions and phases have, in turn,  connections with  the co-flow-only and counter-flow-only phases  in multicomponent superconductors, 
where they can be caused by inter-component electromagnetic coupling
\cite{babaev201547phase,babaev2004superconductor,smiseth2005field,kuklov2006deconfined,herland2010phase}.
Also, the fact that the drag interaction results in an interaction between topological
excitations in different sectors of the model, connects this 
problem to the more general problem of phase transitions in multicomponent gauge theories
\cite{senthil2004deconfined,kuklov2008deconfined,kuklov2006deconfined,herland2013phase,chen2013deconfined,sellin2016first}.

Although the Andreev-Bashkin drag effect is widely expected to be a quite generic and important phenomena in multispecies systems, the magnitude of the drag interaction and its relation to microscopic parameters  was studied only in some special cases. Apart from the rather extensive studies in the context of Fermi liquids in dense nuclear matter
\cite{SJOBERG1976511,PhysRevC.54.2745,BALDO1992349,Chamel} most of the
previous studies include analytic treatment for weakly interacting systems where the effect is inherently weak \cite{fil2005nondissipative,linder2009calculation}, as well as Monte Carlo simulation of zero-temperature $J$-current analog of a two-species Bose-Hubbard model \cite{kaurov2005drag}, mean-field treatment \cite{linder2009calculation}, plane-wave expansion \cite{hofer2012superfluid} and diffusion quantum Monte Carlo \cite{10.1088/1367-2630/aa93a0}. In this article we   calculate the drag  strength dependence on microscopic parameters in the two-species Bose-Hubbard model by means of quantum Monte Carlo.

We now outline how superfluid-superfluid interactions are characterized on an effective field theory level. Superfluidity can be understood in terms of a complex field $\psi=\sqrt{n} e\sp{i\theta}$.
The  kinetic free-energy density of the superflow is
$f=\rho \boldsymbol{v}\sp 2/2$,
where $\boldsymbol{v}=\gamma\nabla\theta$ with 
$\gamma=\hbar/m$ is the superfluid velocity and $\rho=n/\gamma$ is the   superfluid density \cite{fisher1973helicity,svistunov2015superfluid}. It was suggested by Andreev and Bashkin \cite{andreev1976three} that for an interacting binary system (such as superfluid currents in He$^3$/He$^4$-mixtures), a crossterm $\boldsymbol{v}_a\cdot\boldsymbol{v}_b$ is necessarily included to the free energy, where $a$ and $b$ are labels for the two components, with $\boldsymbol{v}_a=\gamma_a\nabla\theta_a$ and $\boldsymbol{v}_b=\gamma_b\nabla\theta_b$. The free-energy density for a two-component superfluid can be written \cite{andreev1976three} \footnote{In some cases, e.g. \cite{fil2005nondissipative,svistunov2015superfluid}, \eqref{eq:Etwospecies} is written on the form $f=(\rho_a-\rho_{ab}) \boldsymbol{v}_a^2/2
	+(\rho_b-\rho_{ab})\boldsymbol{v}_b^2/2
	+\rho_{ab}\boldsymbol{v}_a\cdot\boldsymbol{v}_b$, we however find it more convenient by working in the form of \eqref{eq:Etwospecies} via the redefinitions $\rho_{a/b}-\rho_{ab}\rightarrow\rho_{a/b}$.}
\begin{equation}
	f=\frac{1}{2}\rho_a \boldsymbol{v}_a^2
	+\frac{1}{2}\rho_b\boldsymbol{v}_b^2
	+\rho_{ab}\boldsymbol{v}_a\cdot\boldsymbol{v}_b,
	\label{eq:Etwospecies}
\end{equation}
where
the last term is the drag interaction. The parameter $\rho_{ab}$, the main focus of this article, can be positive or negative. Since $\rho_a$ and $\rho_b$ are positive it follows that $\rho_{ab}^2<\rho_a\rho_b$ in order for  \eqref{eq:Etwospecies} to be bounded from below.
 The mass flow currents $\boldsymbol{j}$ are obtained by differentiation of \eqref{eq:Etwospecies} with respect to the velocities $\boldsymbol{v}$, which gives
\begin{subequations}
\begin{align}
	\boldsymbol{j}_{a}
	&=\rho_{a}\boldsymbol{v}_{a}
	+\rho_{ab}\boldsymbol{v}_{b}\\
	\boldsymbol{j}_{b}
	&=\rho_{b}\boldsymbol{v}_{b}
	+\rho_{ab}\boldsymbol{v}_{a}.
\end{align}\label{eq:massflowcurrents}%
\end{subequations}
Here, the effect of the drag interaction term $\rho_{ab}$ can be seen explicitly: a part of the superflow $\boldsymbol{j}_a$ is due to a \textit{drag} from component $b$, and vice versa. 

Key properties of superfluids depend on the physics of   quantum vortices \cite{Onsager1949,FEYNMAN195517}, i.e. points in space around which the phase $\theta$ \textit{winds} by $2\pi$ so that around a contour $C$ enclosing the vortex it holds that $\oint_C d\boldsymbol{r}\cdot \nabla \theta=2\pi$. Since there are multiple phase fields $\theta$ in multicomponent superfluids, there are multiple types of vortices. The drag effect has important consequences for the interaction of such vortices, since intercomponent vortex interactions are mediated by the drag effect. This is most easily seen by rewriting the free energy \eqref{eq:Etwospecies} in terms of a sum and a difference of the phase gradients \cite{PhysRevB.78.144510}, which gives
\begin{align}
	f=\frac{1}{2}&\left(\frac{\rho_a}{m_a^2}+\frac{\rho_b}{m_b^2}-\frac{2\rho_{ab}}{m_am_b}\right)^{-1}\times\nonumber\\
	\times
	\Bigg[&
	\left(
	\left(\frac{\rho_{a}}{m_a^2}-\frac{\rho_{ab}}{m_am_b}\right)\nabla\theta_a
	-\left(\frac{\rho_b}{m_b^2}-\frac{\rho_{ab}}{m_am_b}\right)\nabla\theta_b\right)^2
	\nonumber
	\\
	&+\frac{\rho_a\rho_b-\rho_{ab}^2}{m_a^2m_b^2}
	\Big(\nabla\theta_a+\nabla\theta_b\Big)^2
	\Bigg],
	\label{eq:separated}
\end{align}
and for the special case with $m_a=m_b=1$ and $\rho_a=\rho_b=\rho$, Eq. \eqref{eq:separated} simplifies to 
\begin{equation}
	f=\frac{\rho-\rho_{ab}}{4}
	\left[\nabla(\theta_a-\theta_b)\right]^2
	+\frac{\rho+\rho_{ab}}{4}
	\left[\nabla(\theta_a+\theta_b)\right]^2.
	\label{eq:illuminating}
\end{equation}
The forms of Eqs. \eqref{eq:separated}, \eqref{eq:illuminating} are particularly illuminating since it is clear how vortices from different components are coupled through their phase windings. It is possible to derive vortex interaction potentials from \eqref{eq:illuminating}. Denoting for a pair of individual vortex lines with cores separated by a distance $r$ and with phase windings $(M_1,M_2)=(\oint_{C_a}d\boldsymbol{r}\cdot\nabla\theta_a/2\pi,\oint_{C_b}d\boldsymbol{r}\cdot\nabla\theta_b/2\pi)$ we have for example for $(1,0)$- and $(0,1)$-vortices the interaction energies 
\begin{subequations}
\begin{align}
	E^{\text{int}}_{(1,0)+(1,0)}&=-\pi\rho\log(r)\\
	E^{\text{int}}_{(1,0)+(0,1)}&=-\pi\rho_{ab}\log(r),
\end{align}
\end{subequations}
where $r$ is the distance between the vortices. Vortices from same component repel logarithmically, but since $\rho_{ab}$ can be either positive or negative vortices from different components can either repel or attract logarithmically. Since the  vortex excitations determine basic properties of superfluids such as rotational response, phase transitions  etc., it is thus clear that the sign and magnitude of $\rho_{ab}$ has crucial effects on the physics of multicomponent systems. 

\section{Model \& background}
In this article we will consider a two-species Bose-Hubbard (BH) model with intraspecies Hamiltonians $H_a$ and $H_b$ containing the on-site intraspecies interactions $U_{a}$ and $U_{b}$, hopping terms $t_{a}$ and $t_{b}$, and chemical potentials $\mu_{a}$ and $\mu_{b}$. The interspecies coupling is through an on-site interaction $U_{ab}$ (which is tunable in experiments \cite{papp2008tunable,thalhammer2008double}), so the Hamiltonian is given by
\begin{align}
	H=H_a+H_b+
	U_{ab}\sum_{i}
	n_{i}^{(a)}n_{i}^{(b)},
	\label{eq:fullHamiltonian}
\end{align}
with the intraspecies Hamiltonians given by 
\begin{align}
	H_a=&-t_{a}\sum_{\langle ij\rangle}
		a_{i}\sp{\dagger(a)} a_{j}\sp{(a)}
	+\frac{U_{a}}{2}\sum_{i}
		n_{i}^{(a)}
		(n_{i}^{(a)}-1)
	\nonumber	
	\\
	&-
	\mu_a\sum_i n_{i}^{(a)},
	\label{eq:intraspeciesHamiltonian}
\end{align}
where $n^{(a)}_i=a_{i}\sp{\dagger(a)} a_{i}\sp{(a)}$ is the number of particles of type $a$ on site $i$.

 We consider here a two-dimensional square lattice with periodic boundary conditions. Eq. \eqref{eq:intraspeciesHamiltonian}  can be obtained as a space-discretized version of the second quantized field description of bosons with truncated interactions \cite{gersch1963quantum}, although as previously mentioned the Hamiltonian was  realized in an optical lattice where the energy scales are experimentally tunable \cite{jaksch1998cold,greiner2002quantum,jaksch2005cold}. The phase diagram of the  one-species Bose-Hubbard model has been studied using various techniques \citep{freericks1994phase,damski2006mott,ejima2012characterization,fisher1989boson,batrouni1992world,prokof1998worm,prokof1998exact,capogrosso2008monte}. The double-species case has been studied experimentally \cite{catani2008degenerate,thalhammer2008double} and with quantum Monte Carlo simulations \cite{PhysRevB.77.184505,soyler2009sign,lingua2015demixing,guglielmino2010mott}. 

\section{Numerical methods}
The BH Hamiltonian can be simulated using Monte Carlo sampling of configurations in the imaginary-time path integral picture \cite{feynman1953atomic,feynman2010quantum,ceperley1995path} with worm \cite{prokof1998worm,prokof1998exact,svistunov2015superfluid} updates. For details, refer to Appendix \ref{appendix:numerics}. In the path-integral picture the system is mapped onto a $(d+1)$-dimensional system, where the extra dimension is imaginary time, and the configuration space becomes a set of string-like structures called worldlines, which are periodic as the imaginary time goes from 0 to $\beta$. If the system is also periodic in the spatial dimensions, some of these ``strings" may, when varying the time from 0 to $\beta$, {\it wind} across a periodic boundary in space. This is the cause of superfluidity, which is measurable with the celebrated Pollock-Ceperley formula \cite{pollock1987path}, which gives the superfluid density $\rho$ for an isotropic system of particles with inverse mass $\gamma=\hbar/m$, volume $L\sp d$ and inverse temperature $\beta$ as
\begin{equation}	
	\rho
	=
	\frac{L^{2-d}}{d\beta}
	\frac{\langle W\sp2 \rangle}{\gamma\sp2},
	\label{eq:Ceperley-Pollock}
\end{equation}
where $\langle W\sp2\rangle$ is the average of the squared \textit{winding number} vector $W\sp2=\sum_i W_i\sp2$. The winding number $W_i=[\boldsymbol{W}]_i$ is the net number of times the particles cross a periodic boundary in the direction $i$  \footnote{To avoid confusion, we point out that this winding number is unrelated to the previously mentioned phase winding of vortices}. The formula \eqref{eq:Ceperley-Pollock} shows that the more the worldlines are disordered, the more the superfluid is ordered.
We are concerned with how quantities like \eqref{eq:Ceperley-Pollock} compare with each other, so the prefactor $L^{2-d}/d\beta$ is unimportant for our purposes. The winding number $W_i=n_{\rightarrow,i}-n_{\leftarrow,i}$ can be straightforwardly calculated in a quantum Monte Carlo simulation by counting the number of times $n_{\rightarrow,i}$ ($n_{\leftarrow,i}$)  a particle crosses a plane perpendicular to the $i$-direction along the positive (negative) direction.

We measure the Andreev-Bashkin interaction by keeping track of the winding numbers for two species $a,b$ and calculate the corresponding densities, $\rho_{a/b}$ from $\langle W_{a/b}\sp2 \rangle$ and $\rho_{ab}$ from $\langle \boldsymbol{W}_{a}\cdot\boldsymbol{W}_{b}\rangle$. A derivation in Appendix \ref{appendix:derivation} gives that the free-energy density for the two-species case is given by
\begin{equation}
	f=\frac{L\sp{2-d}}{\beta d}
	\left(\frac{1}{2}\frac{\langle W_a\sp2\rangle}{\gamma_a\sp2}\boldsymbol{v}_a\sp2
	+\frac{1}{2}\frac{\langle W_b\sp2\rangle}{\gamma_b\sp2}\boldsymbol{v}_b\sp2
	+
	\frac{\langle \boldsymbol{W}_a\cdot\boldsymbol{W}_b\rangle}
	{\gamma_a\gamma_b}\boldsymbol{v}_a\cdot\boldsymbol{v}_b
	\right),
\end{equation}
to be compared with \eqref{eq:Etwospecies}.
We can thus represent $\rho_{ab}$ in terms of the other prefactors $\rho_a$ and $\rho_b$ in \eqref{eq:Etwospecies} with
\begin{equation}
	\varrho_{ab}\equiv\frac{\rho_{ab}}{\sqrt{\rho_a\rho_b}}
	=\frac{
		\langle 
			\boldsymbol{W}_a
			\cdot 
			\boldsymbol{W}_b 
		\rangle
	}
	{
	\sqrt{
		\langle W_a\sp2 \rangle
		\langle W_b\sp2 \rangle
		}
	}.
	\label{eq:rho_ab}
\end{equation}
Clearly, $\varrho_{ab}$ can be either positive or negative, and can not exceed unity in this model.

\section{Results}

\begin{figure}
	\includegraphics[width=\columnwidth]{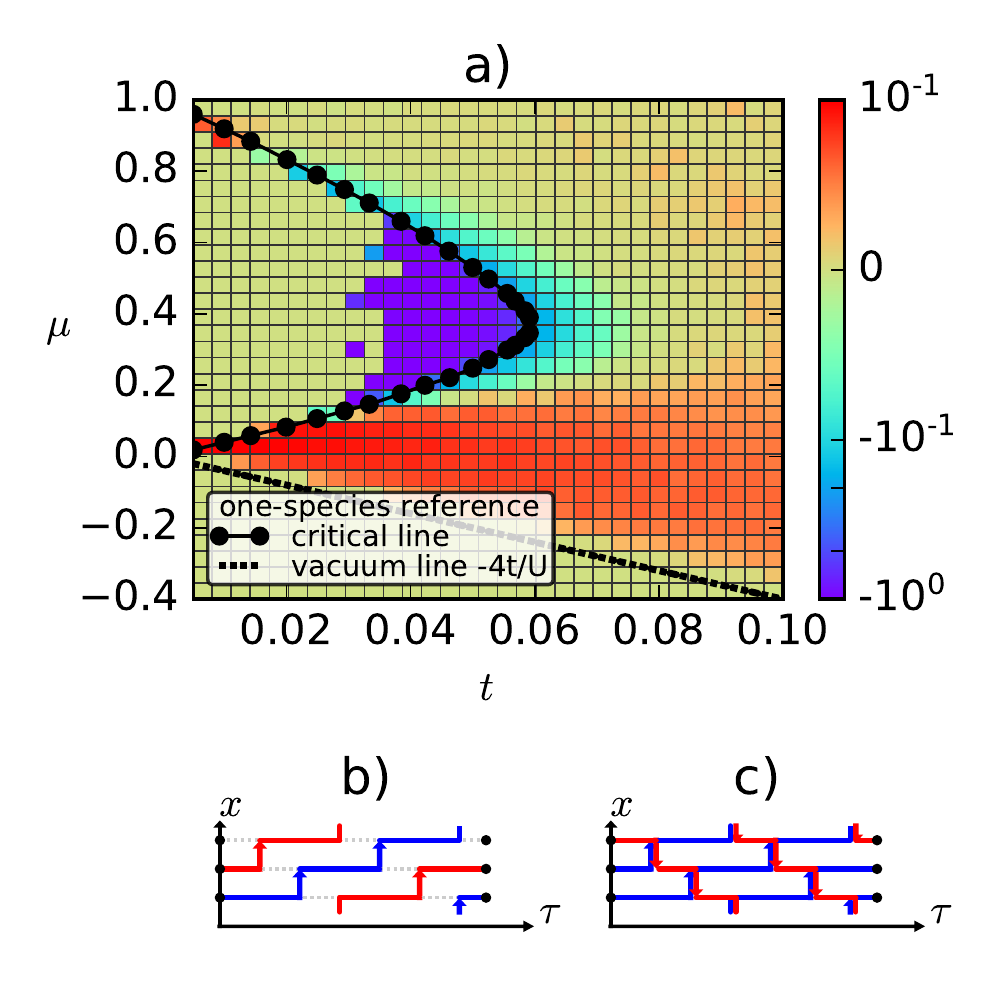}
	\caption{a) The value of the drag interaction $\varrho_{ab}$ for a system with $U_{ab}=0.9$, $L=10$ and $\beta=L/t$, versus chemical potential $\mu$ and hopping parameter $t$. Note that the colormap is logarithmic. For comparison against the one-species phase diagram case we include the one-species critical line (separating the Mott insulating and superfluid phases) which is adapted from Fig. 1. of Ref. \cite{capogrosso2008monte}. In b) and c) we give cartoon illustrations of worldline configurations (in imaginary time $\tau$ and in one spatial dimension $x$) that give a positive and negative drag respectively. In b) the system can avoid paying interspecies interaction energy by having codirected paths, and this effect is strongest in the strongly correlated regime (small $t$) with small filling factor, and in c) a particle-particle counterflow (or a particle-hole co-flow) can be seen.}
	\label{fig:phasediag}
\end{figure}

\begin{figure}
	\includegraphics[width=\columnwidth]{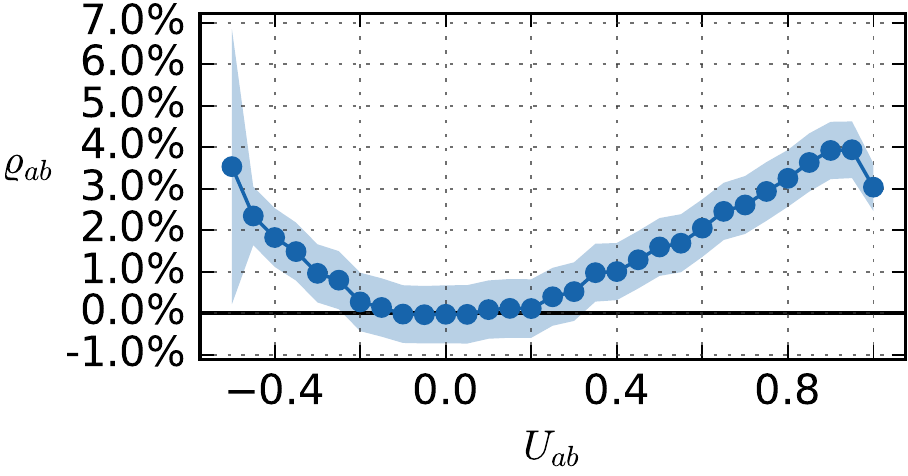}
	\caption{The drag interaction can be induced by a negative or positive intercomponent interaction $U_{ab}$, shown here for the regime where $t=0.1$, $\beta=100$ and $L=10$ and $\mu$ tuned to give an average total number of particles $\langle n_{a+b} \rangle\approx 0.5$.
	}
	\label{fig:mu_vs_Uab}
\end{figure}

\begin{figure}
	\includegraphics[width=\columnwidth]{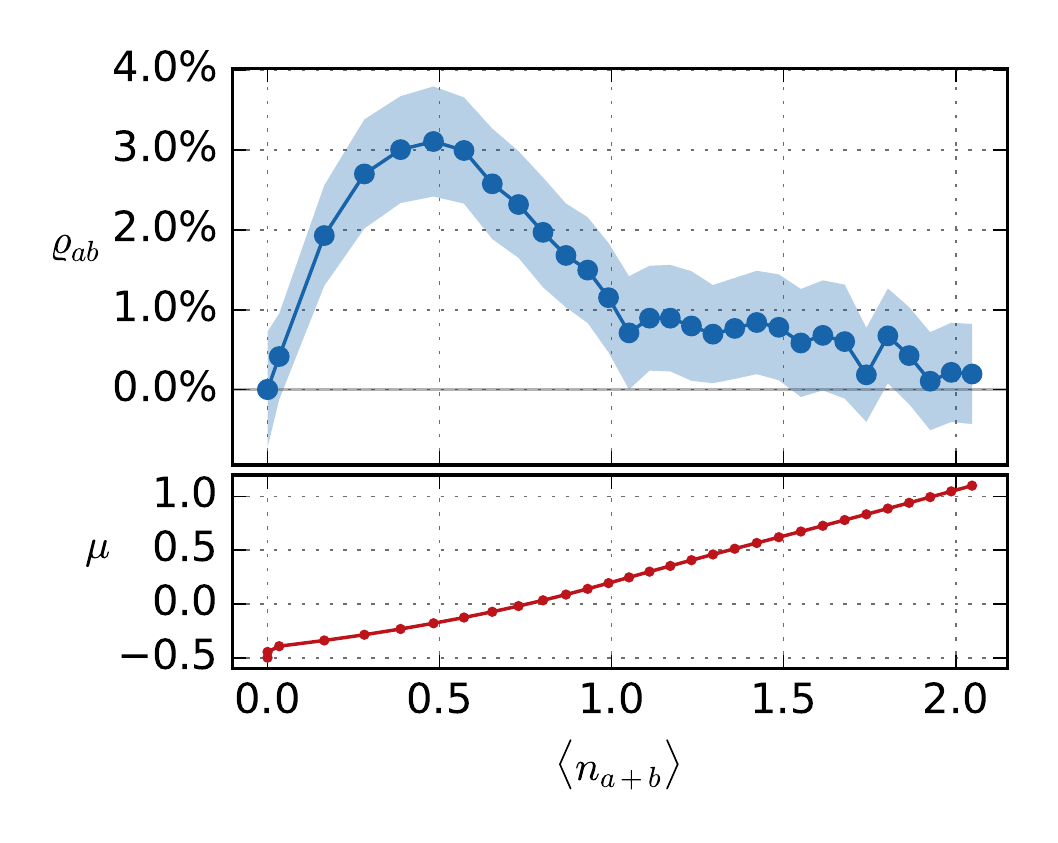}
	\caption{For the case of a positive drag interaction and for parameters
	$U_{ab}=0.8$, $t=0.1$, $L=10$, $\beta=100.0$, we find that it is the largest for low filling factors. In the upper plot we show the superfluid drag interaction versus average filling factor, showing a maximum at around $\langle n_{a+b}\rangle=0.5$. In the lower plot the values of $\mu$ used to set $\langle n_{a+b}\rangle$ is shown.}
	\label{fig:rhoab_vs_nab}
\end{figure}

\begin{figure}
	\includegraphics[width=\columnwidth]{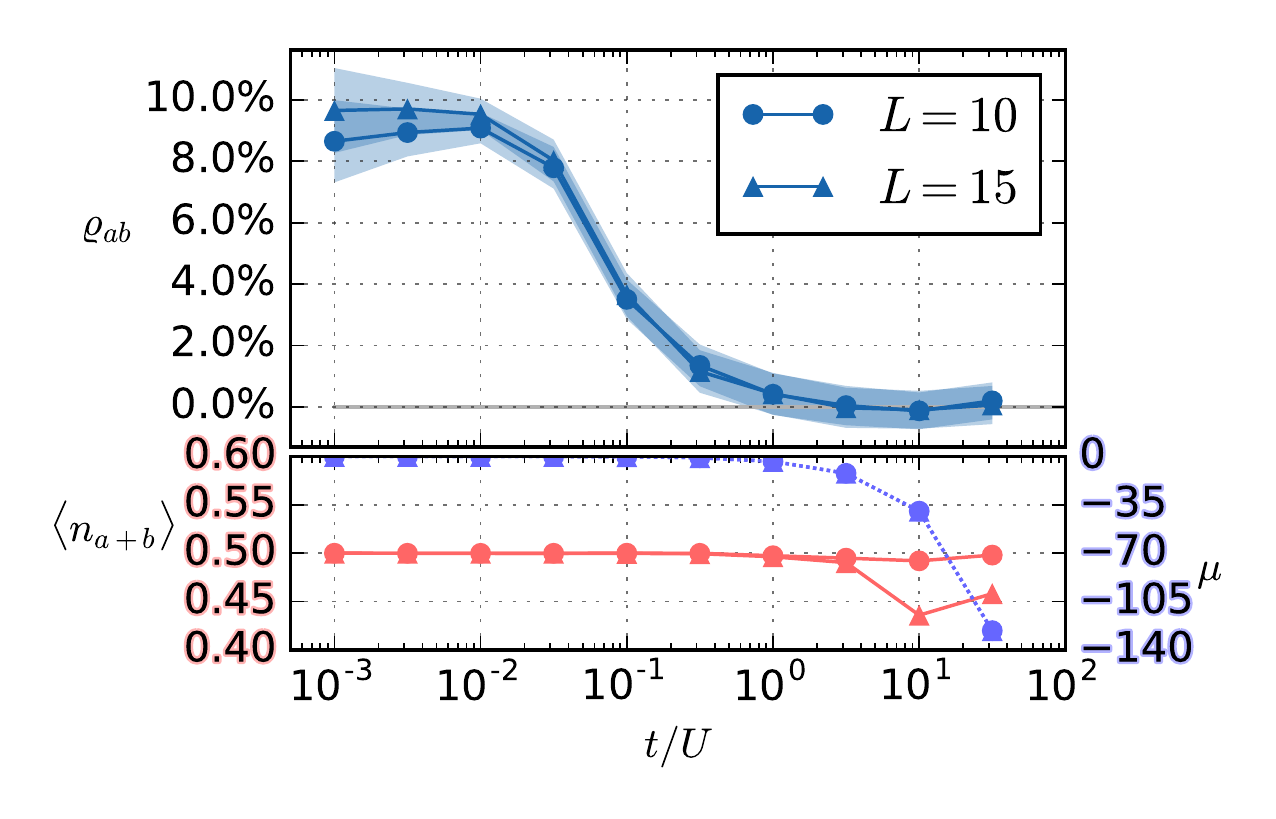}
	\caption{For $U_{ab}=0.9$ and $\beta=L/t$ we find that the drag interaction is relatively large in the strongly-correlated regime with small $t$ and low filling factor. In the upper plot we plot the drag interaction versus $t$ and in the lower plot the value of $\langle n_{a+b}\rangle$ (left axis), and values of $\mu$ used in the simulation (right axis). For the two system sizes simulated we find no substantial finite-size effects.}
	\label{fig:rhoab_vs_tdU}
\end{figure}

\begin{figure}
	\includegraphics[width=\columnwidth]{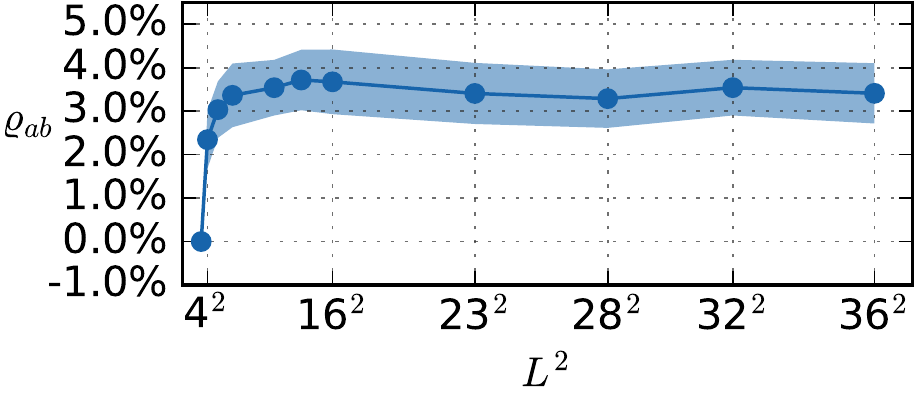}
	\caption{Finite-size scaling analysis of the drag interaction  in the double-superfluid phase shows a saturation at around $L=8$, for a system with $t=0.1$, $\beta=100.0$, $U_{ab}=0.9$ and $\mu=-0.158$.}	
	\label{fig:scaling}
\end{figure}

We now present our results. First we consider the symmetric case with $U_a=U_b=U=1$, $\mu_a=\mu_b=\mu$ and $t_a=t_b=t$. In Fig. \ref{fig:phasediag} we show the drag interaction versus $\mu$ and $t$ for a strong interspecies interaction $U_{ab}=0.9$. We note in Fig. \ref{fig:phasediag} two separate phases in terms of the drag interaction: a double-superfluid phase and a supercounterfluid phase. In the double-superfluid phase of Fig. \ref{fig:phasediag}, the drag interaction is typically a few percent, and can be large and negative close to the supercounterfluid phase. In the supercounterfluid phase \cite{kuklov2004superfluid} the drag interaction is saturated at $\varrho_{ab}=-100\%$. The supercounterfluid state 
is characterized by $\langle (W_a-W_b)^2 \rangle\neq 0$  and $\langle (W_a+W_b)^2 \rangle\rightarrow0$ in the thermodynamic limit
 \cite{soyler2009sign,lingua2015demixing}. For the parts of the double-superfluid phase
 where the filling factor is low and interactions are strong, the system tends to avoid paying interspecies interaction energy by having co-directed paths as shown in Fig. \ref{fig:phasediag} b), giving a positive drag interaction. A strong negative drag is found for the part of the phase diagram where the system would be in a Mott insulating state for the one-species case, however for the two-species case supercounterfluidity occurs via counterdirected paths as illustrated in Fig. \ref{fig:phasediag} c), giving a negative drag interaction.

\subsection{Effect of varying interspecies with low filling factor and $t=0.1$}
We now study how the drag interaction depends on $U_{ab}$ when the mean particle number is fixed. 
We determine the values $\mu$ needed to fix $\langle n_{a+b}\rangle=0.5$, where $n_{a+b}=n_a+n_b$ is the total number of particles per site. We perform grand-canonical simulations allowing particle number fluctuations, but $\mu$ fixes the mean particle number within a few percent using $\mu=-0.297+0.248\cdot U_{ab}-0.108\cdot U_{ab}^2$, which is obtained by running several simulations for different $\mu$, making an linear interpolation fit and solving for $\langle n_{a+b}\rangle=0.5$. In Fig. \ref{fig:mu_vs_Uab} we plot the drag interaction $\varrho_{ab}=\rho_{ab}/\sqrt{\rho_a\rho_b}$ versus $U_{ab}$. In the figure, shaded regions correspond to $1\sigma$ errors estimated by bootstrapping and lines are guides to the eye, this applies also to the other figures in this article. Fig. \ref{fig:mu_vs_Uab}  shows  clearly that there is a drag that is induced by the interspecies interaction $U_{ab}$, and for the systems considered here its magnitude ranges around a few percent of the one-species superfluid densities. 
Note how an attractive interaction and a repulsive interaction both result in a positive drag coefficient in the case of Fig. \ref{fig:mu_vs_Uab}.
This is natural when thinking in terms of the winding of the particle paths in imaginary time, Fig. \ref{fig:phasediag} a), in the repulsive case the paths will tend to not cross, and in the attractive case the paths will tend to superimpose, either case leads to codirected paths.

\begin{figure}	
	\includegraphics[width=\columnwidth]{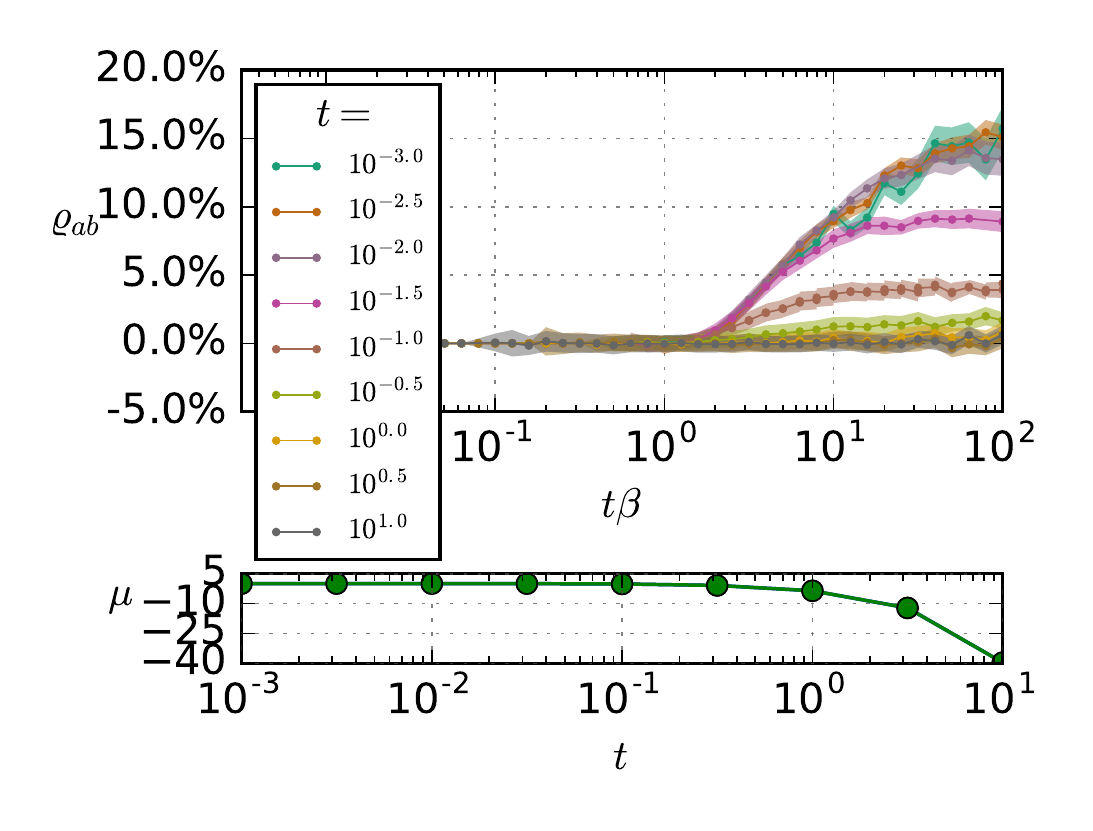}
	\caption{The drag interaction  becomes larger and shows tendency to saturate when increasing the inverse temperature $\beta$, a system with $L=10$ and $U_{ab}=0.9$. In the upper plot we show the drag interaction versus inverse temperature $\beta$, and various $t$, showing a maximum of $\varrho_{ab}=15\%$. In the lower plot we show the chemical potentials used to set $\langle n_{a+b}\rangle=0.5$ for low temperatures.}
	\label{fig:rhoab_vs_beta}
\end{figure}

\begin{figure}	
	\includegraphics[width=\columnwidth]{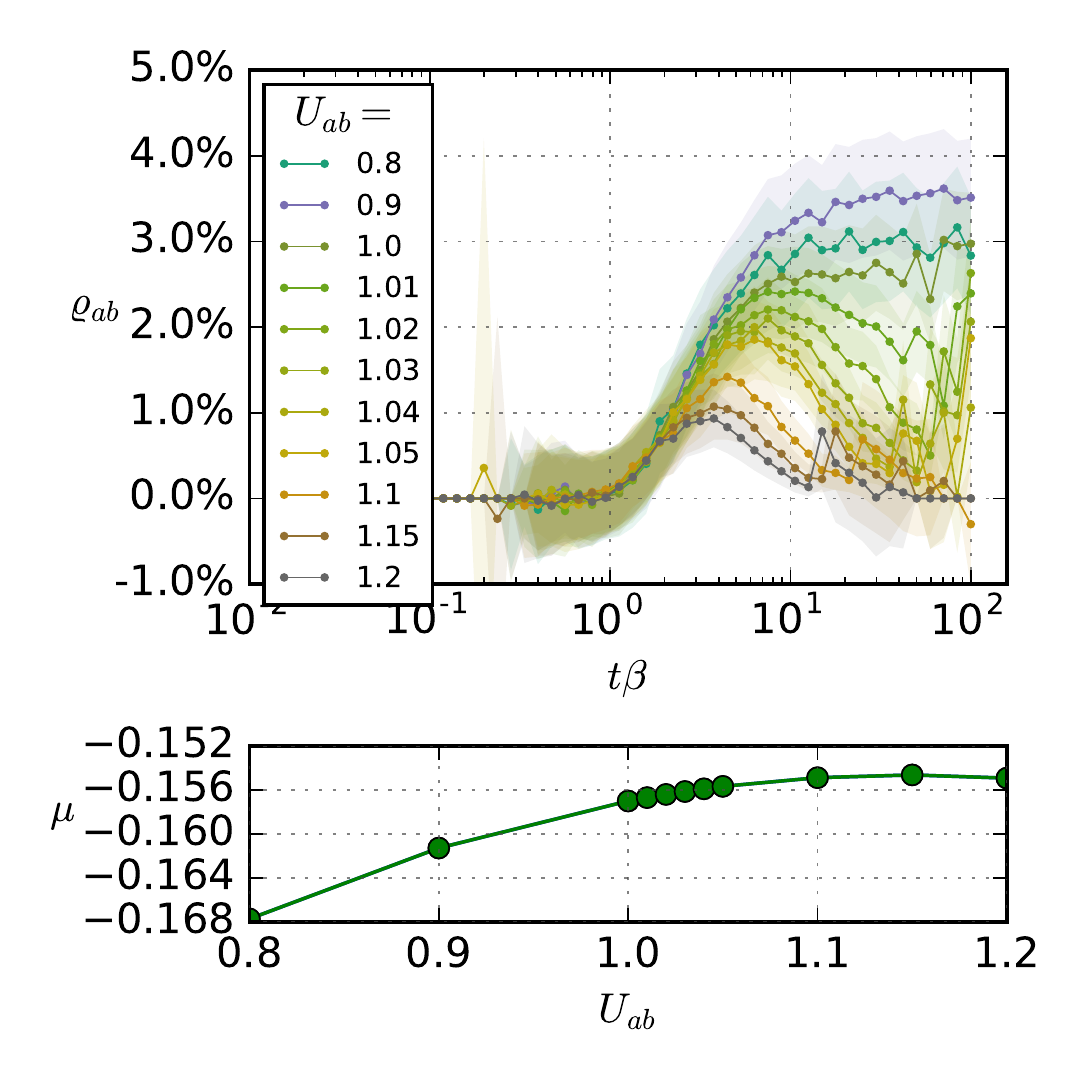}
	\caption{For several systems with values of $U_{ab}$ which gives rise to phase separation in the ground state, we find that the the drag interaction has a non-monotonic dependence on temperature. In the upper plot we show the drag interaction versus inverse temperature $\beta$ for several values of $U_{ab}$. For $U_{ab}\geq1$, the system is phase separated for small temperatures. Increasing the temperature leads to thermal mixing and the system thus goes to the double-superfluid regime and acquires an intercomponent drag.  Here $L=6$, and $t=0.1$. In the lower plot we show the values of $\mu$ used to give $\langle n_{a+b}\rangle=0.5$ for the lowest temperatures.}
	\label{fig:rhoab_vs_beta_phase_sep}
\end{figure}

\subsection{Varying filling factor and $t$}
We now turn to the dependence of the drag interaction to the lattice filling. In Fig. \ref{fig:rhoab_vs_nab} we fix $U_{ab}=0.8$, $t=0.1$ and vary $\mu$, thereby varying the total filling factor $\langle n_{a+b}\rangle$. As is seen, the drag interaction has a maximum around $\langle n_{a+b}\rangle=0.5$. For low densities the worldline paths can avoid each other to minimize the interaction energy stemming from $U_{ab}$, leading to correlations between the winding numbers and thus an effective interaction. For larger densities, this way of minimizing the interaction energy is not effective, the only way would be to phase-separate the system. Note that for other values of $t$ these results may look quite different, as the system enters a supercounterfluid phase for lower $t$, see Fig. \ref{fig:phasediag}. Next, lets consider effects of varying $t$, in Fig. \ref{fig:rhoab_vs_tdU} it is seen that the drag interaction drops in the weak-coupling regime of large $t$. 

\begin{figure}	
	\includegraphics[width=\columnwidth]{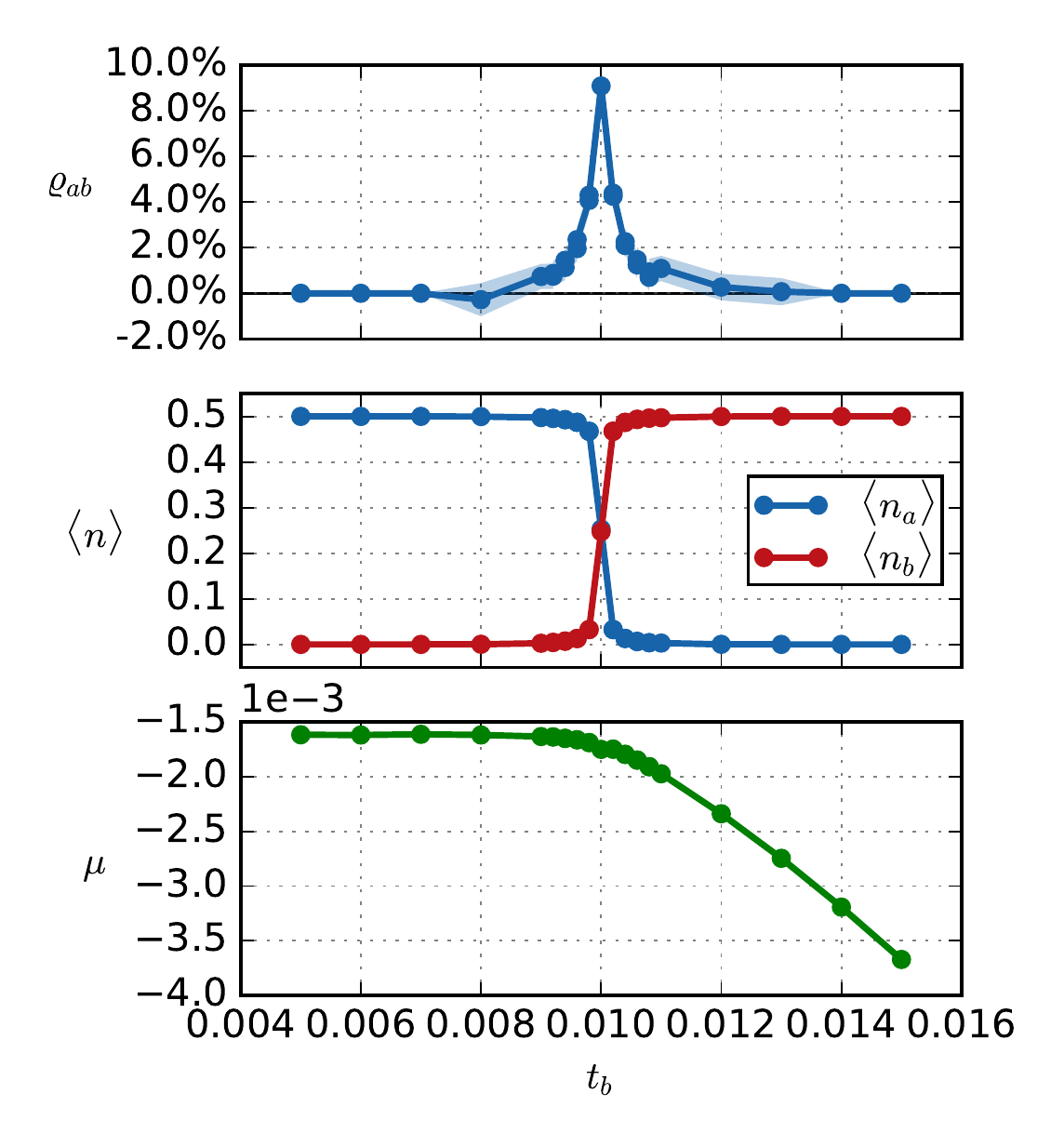}
	\caption{For a system with $t_a=0.01$ is fixed and $t_b$ varies, $U_{ab}=0.9$, $\beta=1000$, $L=10$, and $\mu=\mu_a=\mu_b$ is tuned set $\langle n_{a+b}\rangle=0.5$, we find that the system phase separates. The drag is the largest near the phase separation phase transition  and the  disappears in the phase separated regime.}
	\label{fig:rhoab_vs_mubdUb}
\end{figure}

\subsection{Finite-size scaling in the double-superfluid phase} 
Since the physics considered so far are away from criticality we do not expect them to be altered much by finite-size scaling beyond a saturation point. We have performed a finite-size scaling analysis in Fig. \ref{fig:scaling} where we plot $\varrho_{ab}$ for various system sizes, for parameters corresponding to the system in Fig. \ref{fig:mu_vs_Uab}, in the double-superfluid phase.
In Fig. \ref{fig:scaling} we show how the drag interaction scales with $L^2$, showing a quick saturation when scaling up the system size and suggests that the drag interaction becomes independent of system size for large enough systems in the double-superfluid phase. 

\begin{figure}	
	\includegraphics[width=0.9\columnwidth]{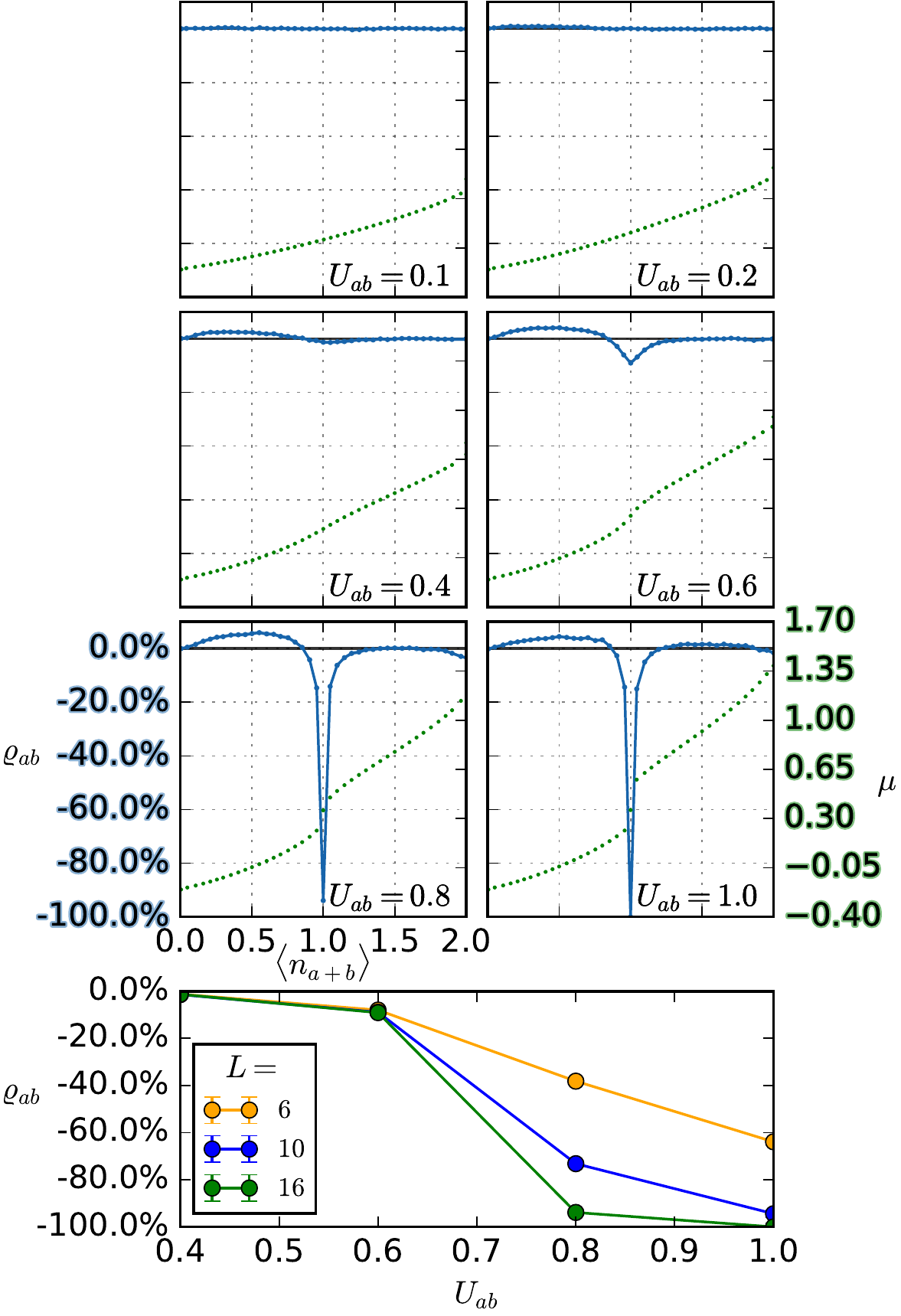}
	\caption{The drag interaction changes drastically when crossing the border between a double-superfluid phase with an unsaturated drag to a supercounterfluid phase where the drag is saturated at $-100\%$. Here we display for various $U_{ab}$, one in each panel for the six upper panels, the dependence of the drag interaction (blue solid lines, left $y$-axes) versus the filling factor $\langle n_{a+b}\rangle$ for $t=0.05$ and $L=16$. The corresponding values for the chemical potentials are given on the right $y$-axes (green dots). It is seen that for low filling factors with $\langle n_{a+b}\rangle<1$, there is a positive drag interaction which becomes larger with increasing intercomponent interaction $U_{ab}$. If the system approaches criticality, for large enough $U_{ab}$ the drag first goes down to zero and then changes its signs to negative (i.e. co-flow drag crosses over to counterflow), then the absolute value of negative drag starts growing rapidly when approaching $\langle n_{a+b}\rangle=1$. When the system enters the supercounterfluid phase the drag saturates to the lower extreme value of $\varrho_{ab}=-100\%$. In the lower panel, we show the value of the drag at $\langle n_{a+b}\rangle=1$, for three system sizes, showing that finite-size effects are important in the supercounterfluid regime.} 
	\label{fig:SCF}
\end{figure}

\subsection{Temperature dependence of drag and emergence of a drag interaction in systems with phase separated ground-state}

Next we consider the effects of changing temperature. In Fig. \ref{fig:rhoab_vs_beta} we display the dependance of $\rho_{ab}$ on the inverse tempature $\beta$ {(for small $\beta$ data points where $\rho_{ab}/\sqrt{\rho_a\rho_b}=0/0$ are obtained, these has been represented as 0 in Fig. \ref{fig:rhoab_vs_beta})}. The drag interaction is seen in Fig. \ref{fig:rhoab_vs_beta} to decrease with temperature. The same trend is seen  analytical studies of weakly coupled systems, Fig. 1 of \cite{fil2005nondissipative}. 

 However, the drag effect can indeed have more complicated temperature dependence.
For a sufficiently strong intercomponent interaction $U_{ab}$, the system can phase separate in the ground state. In a phase separated system there is clearly no drag interaction. However, with increasing temperature thermal fluctuations in form of a remixing with the decimated particle kind should occur, inducing a drag interaction, which should then disappear as superfluidity is destroyed by further increasing temperature. In Fig. \ref{fig:rhoab_vs_beta_phase_sep} we can observe precisely this effect, where the drag interaction emerges at elevated temperatures. Overall it depends non-monotonically on temperature for systems which are phase-separated in the ground state. The drag should in fact  have a broader relevance  for phase separated systems because  the system can have  a rotation-induced remixing even at low temperatures, as follows from Gross-Pitaevskii-model-based studies \cite{galteland2015thermal}. 

\subsection{Effects of density imbalance}
Until now we have considered cases with symmetric system parameters $U_a=U_b=1$, $t_a=t_b$ and $\mu_a=\mu_b$. We now also consider the effects of model parameter assymetries, such as for a $^{87}$Rb/$^{41}$K mixture \cite{thalhammer2008double}. We let $U_a=U_b=1.0$, $U_{ab}=0.9$, $t_a=0.01$, $\beta=1000$ and vary $t_b$. To fix $\langle n_{a+b}\rangle$ we also vary $\mu_a=\mu_b=\mu$. The results are shown in Fig. \ref{fig:rhoab_vs_mubdUb} and in this cases the drag interaction effect diminishes with increasing model parameter asymmetry. Note that the mean number of particles of species $a$ and $b$ are not necessarily equal anymore.  

\subsection{Drag near phase transitions to supercounterfluid and paired states}
As was seen in Fig. \ref{fig:phasediag}, the drag interactions changes drastically while going from the double-superfluid phase  with an unsaturated drag to the supercounterfluid phase with a saturated drag of $-100\%$. For $t=0.05$ and at unity filling $\langle n_{a+b}\rangle=1$, this transition was found in a finite-size scaling analysis \cite{lingua2015demixing} to occur at $U_{ab}/t=14.9\pm0.1$ (corresponding to $U_{ab}/U=0.74-0.75$). To see more specifically how the drag interaction behaves at the transition we take $t=0.05$ and vary the filling factor for several intercomponent interactions $U_{ab}$. In Fig. \ref{fig:SCF} we show values of $\varrho_{ab}$ versus $\langle n_{a+b}\rangle$ (set by $\mu$) for various $U_{ab}$. As is seen $U_{ab}$ induces an effective drag interaction for densities $\langle n_{a+b}\rangle<1$, where the system is in the double-superfluid phase. For large $U_{ab}$  there is an onset of a saturated  negative drag regime 
around $\langle n_{a+b}\rangle=1$ which signals the phase transition into the supercounterfluid phase. We point out that there are many values of $\mu$ that causes $\langle n_{a+b}\rangle=1$, so the supercounterfluid would be a plateau rather than a point if having $\mu$ as the horizontal axis (compare with Fig. \ref{fig:phasediag}). For the systems in Fig. \ref{fig:SCF}, we considered three system sizes $L=6,10$ and $16$ and found that finite-size effects are within or close to being within statistical error in the double-superfluid phase, whereas they are sizable in the supercounterfluid phase. For a more careful finite-size scaling study of the supercounterfluid phase, see \cite{lingua2015demixing}.

\begin{figure}	
	\includegraphics[width=\columnwidth]{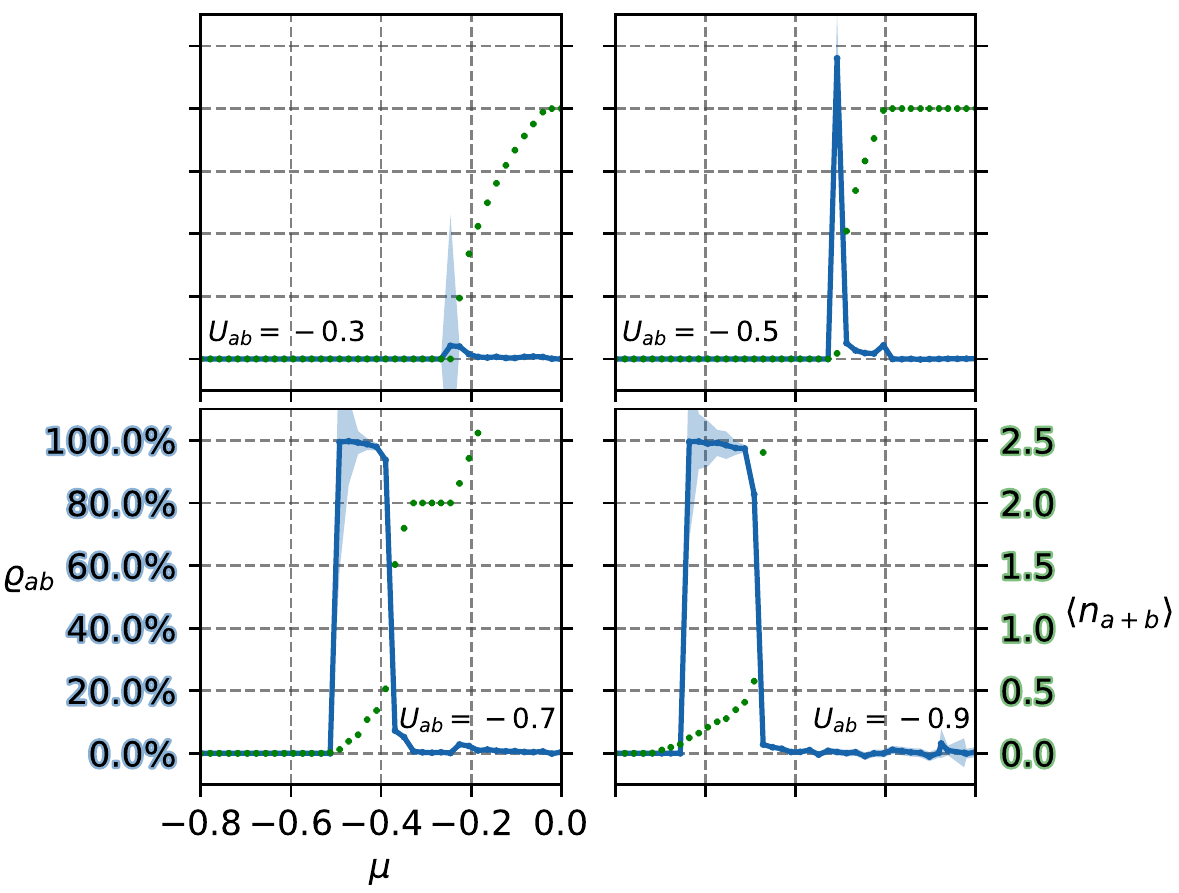}
	\caption{For negative $U_{ab}$, it is possible to have an positive drag that saturates to the upper extreme value of $\varrho_{ab}=+100\%$, in a so called \textit{paired superfluid} phase, as demonstrated here for a system with $t=0.05$, $L=10$, $\beta=200$, and various negative $U_{ab}$. } 
	\label{fig:PSF}
\end{figure}

Until now, we have mentioned the double-superfluid phase with an unsaturated drag interaction, and the supercounterfluid phase with a drag saturated to $-100\%$. In classical loop-current models of two-component bosons, it was demonstrated that there exists also a phase where the drag is saturated to $+100\%$, called the \textit{paired superfluid} phase \cite{kuklov2004commensurate,kuklov2004superfluid,kuklov2006deconfined,herland2010phase},  which appears for a different sign of the interspecies interaction. To check the presence of that phase  in our model, we have performed simulations similar to those of Fig. \ref{fig:SCF}, but with a \textit{negative} interspecies interaction $U_{ab}$.
In Fig. \ref{fig:PSF} we show the dependence of the drag interaction on the chemical potential
for several systems with attractive interspecies interaction. 
 Note that  in Fig. \ref{fig:PSF} the filling factor behaves more abruptly with chemical potential than in Fig.  \ref{fig:SCF}, so it is more convenient to have $\mu$ as the horizontal axis. 
The drag interaction can indeed take the value $+100\%$ for filling factors $0<\langle n_{a+b}\rangle<2$ (which can be compared with the result of Fig. \ref{fig:rhoab_vs_nab}) for the systems considered in Fig. \ref{fig:PSF}. This gives an example of a phase with a drag saturated to $+100\%$. Importantly,  the positive drag could also be expected to be very large in the vicinity of the phase transition to that phase.

\begin{figure}
	\includegraphics[width=\columnwidth]{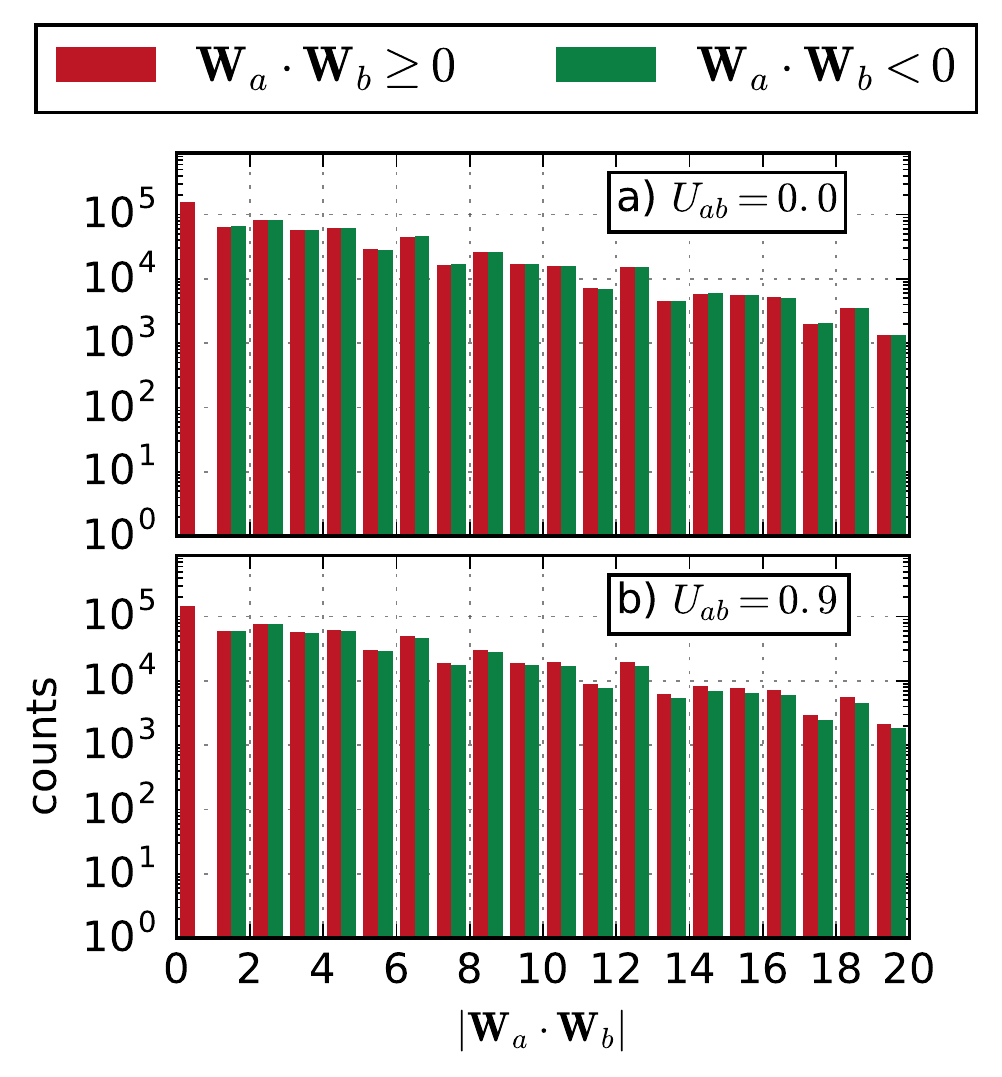}
	\caption{
Histograms of the winding number mixing ${W}_a\cdot{W}_b$ for simulations where the interspecies interaction $U_{ab}$ is a) zero and b) non-zero.  The histograms show an asymmetry for non-zero $U_{ab}$ which means that co-directed paths are slightly more common than counter-directed paths,  leading to an effective positive drag interaction. Other parameters are $U=1$, $t=0.1$, and varying $\mu$ so to have the same $\langle n_{a+b}\rangle$ in a) and b).
}
	\label{fig:histogram_compare_t=0.1}
\end{figure}

\subsection{Statistical nature of the drag interaction}
Finally we  comment on the statistical nature of the drag interaction  in the double-superfluid phase. In Fig. \ref{fig:histogram_compare_t=0.1} we plot histograms for $\boldsymbol{W}_a\cdot \boldsymbol{W}_b$, obtained from simulations for zero and non-zero $U_{ab}$. The values of $\rho_{ab}/\sqrt{\rho_a\rho_b}$ of Fig. \ref{fig:mu_vs_Uab} are calculated using Eq. \eqref{eq:rho_ab} from data sets like the ones depicted in Fig. \ref{fig:histogram_compare_t=0.1}. For $U_{ab}=0$, Fig. \ref{fig:histogram_compare_t=0.1} a), positive and negative values of $\boldsymbol{W}_a\cdot \boldsymbol{W}_b$ are, of course, equally probable since the two species represent two identical and decoupled systems. For a sufficiently large non-zero $U_{ab}$ however, Fig. \ref{fig:histogram_compare_t=0.1}) b), the distribution is instead asymmetric with a higher probability of positive values of $\boldsymbol{W}_a\cdot\boldsymbol{W}_b$. Although the asymmetry between positive and negative values of $\boldsymbol{W}_a\cdot\boldsymbol{W}_b$ in the double-superfluid phase is not necessarily striking, it can as we have seen nevertheless be sufficient to lead to a significant effective interaction. 
If picking out a particular configuration from a simulation in the double-superfluid phase, it is thus not at all certain that one will observe co-directed paths, it is only on a statistical level that the drag interaction emerges \footnote{A remark concerning the co- or counterdirection of superfluid currents is in place to avoid confusion. If the drag interaction $\rho_{ab}$ in \eqref{eq:Etwospecies} is positive, the system can lower its energy by having \textit{ counterdirected} superfluid currents. However, a positive $\rho_{ab}$ corresponds to positive $\boldsymbol{W}_a\cdot\boldsymbol{W}_b$, that is {\it codirected} paths. This is not a contradiction since the imaginary-time paths are not the real superfluid currents, or loosely speaking, the quantity $d \boldsymbol{r}_a / d\tau\cdot d\boldsymbol{r}_b /d\tau$ switches sign when replacing $\tau$ with $i t$.}.

\section{Conclusions}
In conclusion, the intercomponent drag interaction  is a rather generic phenomenon in multicomponent
superfluid and superconducting mixtures.
This interaction is  very important   for phase diagrams, nature  of phase transitions, rotational and magnetic responses and properties of topological defects.
Here we have considered the  origin of the drag in a two-component boson system on a lattice that
have only on-site boson-boson interactions.
We obtained the strength of the intercomponent drag interaction  as a function of microscopic parameters
of the system. 
 The drag  gradually saturates to  $\pm100\%$ close to the phase transitions to paired superfluid and supercounterfluid states, with $+100\%$ drag leading to a complete entrainment of superfluid mass flow \eqref{eq:massflowcurrents}, and $-100\%$ to a complete counterflow. We find that the drag can be substantial even  away from these transitions and even in the case of
on-site intercomponent boson-boson interactions.
The most straightforward experimental visualization of the    drag can be through observation of the structure of vortex lattices, for which symmetry and ordering
  are very sensitive to drag strength \cite{PhysRevB.78.144510,PhysRevLett.101.255301}.

\section*{Acknowledgements}
We thank N. Prokof'ev,  B. Svistunov, D. Weston and E. Blomquist for useful discussions, and A. Kuklov for comments on our manuscript.
The work was supported by the Swedish Research Council Grants No. 642-2013-7837, VR2016-06122 and { G\"oran} Gustafsson Foundation for Research in Natural Sciences and Medicine. 
The computations were performed on resources provided by the Swedish 
National Infrastructure for Computing (SNIC) at National Supercomputer Center in Link\"oping, Sweden. 

\appendix
	
\section{Derivation of a Pollock-Ceperley formula for the two species case}\label{appendix:derivation}
We follow the one-species derivation of Pollock and Ceperley \cite{pollock1987path}, see also \cite{svistunov2015superfluid}, but introduce \textit{two} phase twists $\boldsymbol{\varphi}_\alpha=L\boldsymbol{v}_\alpha/\gamma_\alpha$, with $\alpha=a,b$, and associated winding numbers $\boldsymbol{W}_\alpha$. As detailed in \cite{pollock1987path}, a Galilean transformation (putting the superfluid component in motion) will, due to periodic boundary conditions, multiply the partition function with the factors $\exp(\pm i\boldsymbol{\varphi}\cdot\hat{\boldsymbol{\mu}})$, one for each particle that crosses the periodic boundary in the direction $\pm\hat{\boldsymbol{\mu}}$ (where $\hat{\boldsymbol{\mu}}=\hat{\boldsymbol{x}},\hat{\boldsymbol{y}},\hat{\boldsymbol{z}}$). Defining the winding number $\boldsymbol{W}$ as the net number of times the particles wind across the boundaries, a phase twist then gives in total the factor $\exp(i\boldsymbol{\varphi}\cdot\boldsymbol{W})$ multiplying the partition function. We now proceed by decomposing the partition function in terms of fixed winding-number partition functions
\begin{equation}
	Z_{\boldsymbol{\varphi}_a,\boldsymbol{\varphi}_b\neq\boldsymbol{0}}
	=\sum_{\boldsymbol{W}_a,\boldsymbol{W}_b}
	e^{i(
	\boldsymbol{\varphi}_a\cdot\boldsymbol{W}_a
	+
	\boldsymbol{\varphi}_b\cdot\boldsymbol{W}_b	
	)	
	}Z_{\boldsymbol{W}_a,\boldsymbol{W}_b},
\end{equation}
and by writing the corresponding phase-twist free partition function as
\begin{equation}
	Z_{\boldsymbol{\varphi}_a=\boldsymbol{\varphi}_b=\boldsymbol{0}}
	=\sum_{\boldsymbol{W}_a,\boldsymbol{W}_b}Z_{\boldsymbol{W}_a,\boldsymbol{W}_b}.
\end{equation}
Denoting the free-energy difference associated with introducing the phase twists by $F=F_{\boldsymbol{\varphi}_a,\boldsymbol{\varphi}_b\neq\boldsymbol{0}}-F_{\boldsymbol{\varphi}_a=\boldsymbol{\varphi}_b=\boldsymbol{0}}$, we can write
\begin{align}
	e\sp{-\beta F}&=
	\frac{e\sp{-\beta F_{\boldsymbol{\varphi}_a,\boldsymbol{\varphi}_b\neq\boldsymbol{0}}}}
	{e\sp{-\beta 
	F_{\boldsymbol{\varphi}_a=\boldsymbol{\varphi}_b=\boldsymbol{0}}}}
	=\frac{Z_{\boldsymbol{\varphi}_a,\boldsymbol{\varphi}_b\neq\boldsymbol{0}}}
	{Z_{\boldsymbol{\varphi}_a
	=
	\boldsymbol{\varphi}_b=\boldsymbol{0}}}\\
	&
	=\frac{\sum_{\boldsymbol{W}_a,\boldsymbol{W}_b}
	e^{i
	(\boldsymbol{\varphi}_a\cdot\boldsymbol{W}_a
	+\boldsymbol{\varphi}_b\cdot\boldsymbol{W}_b
	)	
	}Z_{\boldsymbol{W}_a,\boldsymbol{W}_b}}
	{\sum_{\boldsymbol{W}_a,\boldsymbol{W}_b}Z_{\boldsymbol{W}_a,\boldsymbol{W}_b}}\\
	&
	=\langle e\sp{i
	(
	\boldsymbol{\varphi}_a\cdot\boldsymbol{W}_a
	+	\boldsymbol{\varphi}_b\cdot\boldsymbol{W}_b
	)	
	}
	\rangle.\label{right-hand-side}
\end{align}
Expanding the right-hand-side for small phase twists, using that $\langle \boldsymbol{W}\rangle=\boldsymbol{0}$ and assuming a $d$-dimensional system which is isotropic so that winding numbers in different dimensions are uncorrelated, leaves second order terms of the form 
\begin{align}
\langle(\boldsymbol{\varphi}_a\cdot\boldsymbol{W}_a)(\boldsymbol{\varphi}_b\cdot\boldsymbol{W}_b)\rangle
&=
\langle \varphi_{ai}W_{ai}\varphi_{bj}W_{bj}\rangle\\
&=
\langle W_{ai}W_{bj}\rangle\varphi_{ai}\varphi_{bj}\\
&=
\frac{\delta_{ij}\langle \boldsymbol{W}_a\cdot\boldsymbol{W}_b\rangle}{d}\varphi_{ai}\varphi_{bj}\\
&=
\frac{\langle\boldsymbol{W}_a\cdot\boldsymbol{W}_b\rangle}{d}\boldsymbol{\varphi}_a\cdot\boldsymbol{\varphi}_b,
\end{align} 
correspondingly $\langle(\boldsymbol{\varphi}_a\cdot\boldsymbol{W}_a)^2\rangle=\langle W_a^2\rangle\varphi_a^2/d$ and $\langle(\boldsymbol{\varphi}_b\cdot\boldsymbol{W}_b)^2\rangle=\langle W_b^2\rangle\varphi_b^2/d$. Expanding also $e^{-\beta F}$ gives the free-energy density
\begin{equation}
	f=\frac{L\sp{2-d}}{\beta d}
	\left(\frac{1}{2}\frac{\langle W_a\sp2\rangle}{\gamma_a\sp2}\boldsymbol{v}_a\sp2
	+\frac{1}{2}\frac{\langle W_b\sp2\rangle}{\gamma_b\sp2}\boldsymbol{v}_b\sp2
	+
	\frac{\langle \boldsymbol{W}_a\cdot\boldsymbol{W}_b\rangle}
	{\gamma_a\gamma_b}\boldsymbol{v}_a\cdot\boldsymbol{v}_b
	\right).
	\label{binaryquadraticform}
\end{equation}
A generalization to $N$ species follows straightforwardly, as \eqref{binaryquadraticform} is a binary quadratic form the general case is a corresponding $N$-ary quadratic form.

\section{Details of the numerics}\label{appendix:numerics}

The equilibrium statistics of \eqref{eq:fullHamiltonian} can be obtained from quantum Monte Carlo simulations by sampling imaginary-time path integrals. Since we are interested in simulating the superfluid phase, efficient sampling of winding numbers is needed. The winding number is a global topological quantity and consequenctly local Monte Carlo updates will lead to slow convergence of the winding number \cite{pollock1987path}. Here we use the ``worm" algorithm \cite{prokof1998worm,prokof1998exact,svistunov2015superfluid} which efficiently generates configurations with different winding numbers. In short, the worm algorithm is a Metropolis \cite{metropolis1953equation} sampling algorithm  which efficiently generates path integrals by taking shortcuts through an extended configuration space where the particle number is not conserved at two points in spacetime. The points where a particle is created/destroyed are referred to as the tail/head of a worm. The insertion of such discontinuities is referred to as the creation of a worm which is one of the Monte Carlo updates in a worm Monte Carlo simulation. After a worm creation, either one or both of the discontinuities can then sample the extended configuration space by propagating through space-time by a set of local Monte Carlo updates.

The Monte Carlo updates include the aforementioned worm creation update, as well as a time-shift update which displaces a discontinuity in time, a jump update which shifts a discontinuity to a neighboring site thereby inserting a kink, a corresponding anti-jump update which undoes the jump update, a reconnection update which inserts a hole next to the discontinuity, and a correspond anti-reconnection update, and finally, a worm destruction update. Whenever the head and tail of the worm are lined up, meaning that they are in the same point in space and have no events ocurring in between them in time, the worm destruction move may be called which removes the worm and the particle discontinuities from the system. The resulting state then belongs to the ordinary configuration space where particle number is conserved, which counts to the Monte Carlo statistics of the simulation. For a didactic summary of the worm updates, see e.g. \cite{trefzger2011ultracold}. Since the extended configuration space Monte Carlo moves are all detailed balanced, the effective  Monte Carlo moves which updates between ordinary configuration space states sampled are also balanced.  The advantages of the worm algorithm are listed in \cite{prokof1998exact}, for our purposes are the most important ones are that the simulations may sample any winding number, suffer less from critical slowing down, and that it can work in the grand-canonical ensemble.

To simulate a double-species system, we insert worms that operate on particle numbers of one species type one at a time, and may use the same updates as used for a single-species case. The exception is the time-shift update, which for a single-species simulation normally for simplicity updates the time of a worldline discontinuity with the restriction of the nearest lying kinks. To preserve ergodicity in a multi-species simulation the discontinuity must be able to cross kinks that belong to other species. We have performed simulation with $10^8-10^{10}$ worm updates per site and species, after an equilibration warmup of typically $10^8$ updates, and have calculated averages from $10^5-10^6$ data points from each simulation. The initial configurations are ordered states, typically a checkerboard state with $n_i^{(a)}+n_i^{(b)}=1$.
To estimate errors we have used the bootstrap method \cite{barkema2001monte}. We find that around $10^4$ data points are needed in a bin in order to not over-estimate the error for the winding number statistics. The displayed errors for the quantity $\varrho_{ab}=\rho_{ab}/\sqrt{\rho_a\rho_b}$ are the estimated errors of $\rho_{ab}$ divided by $\sqrt{\rho_a\rho_b}$. 

\bibliographystyle{apsrev4-1}
\bibliography{references.bib}

\begin{thebibliography}{70}%
\makeatletter
\providecommand \@ifxundefined [1]{%
 \@ifx{#1\undefined}
}%
\providecommand \@ifnum [1]{%
 \ifnum #1\expandafter \@firstoftwo
 \else \expandafter \@secondoftwo
 \fi
}%
\providecommand \@ifx [1]{%
 \ifx #1\expandafter \@firstoftwo
 \else \expandafter \@secondoftwo
 \fi
}%
\providecommand \natexlab [1]{#1}%
\providecommand \enquote  [1]{``#1''}%
\providecommand \bibnamefont  [1]{#1}%
\providecommand \bibfnamefont [1]{#1}%
\providecommand \citenamefont [1]{#1}%
\providecommand \href@noop [0]{\@secondoftwo}%
\providecommand \href [0]{\begingroup \@sanitize@url \@href}%
\providecommand \@href[1]{\@@startlink{#1}\@@href}%
\providecommand \@@href[1]{\endgroup#1\@@endlink}%
\providecommand \@sanitize@url [0]{\catcode `\\12\catcode `\$12\catcode
  `\&12\catcode `\#12\catcode `\^12\catcode `\_12\catcode `\%12\relax}%
\providecommand \@@startlink[1]{}%
\providecommand \@@endlink[0]{}%
\providecommand \url  [0]{\begingroup\@sanitize@url \@url }%
\providecommand \@url [1]{\endgroup\@href {#1}{\urlprefix }}%
\providecommand \urlprefix  [0]{URL }%
\providecommand \Eprint [0]{\href }%
\providecommand \doibase [0]{http://dx.doi.org/}%
\providecommand \selectlanguage [0]{\@gobble}%
\providecommand \bibinfo  [0]{\@secondoftwo}%
\providecommand \bibfield  [0]{\@secondoftwo}%
\providecommand \translation [1]{[#1]}%
\providecommand \BibitemOpen [0]{}%
\providecommand \bibitemStop [0]{}%
\providecommand \bibitemNoStop [0]{.\EOS\space}%
\providecommand \EOS [0]{\spacefactor3000\relax}%
\providecommand \BibitemShut  [1]{\csname bibitem#1\endcsname}%
\let\auto@bib@innerbib\@empty
\bibitem [{\citenamefont {Andreev}\ and\ \citenamefont
  {Bashkin}(1975)}]{andreev1976three}%
  \BibitemOpen
  \bibfield  {author} {\bibinfo {author} {\bibfnamefont {A.~F.}\ \bibnamefont
  {Andreev}}\ and\ \bibinfo {author} {\bibfnamefont {E.~P.}\ \bibnamefont
  {Bashkin}},\ }\href
  {http://www.jetp.ac.ru/cgi-bin/e/index/e/42/1/p164?a=list} {\bibfield
  {journal} {\bibinfo  {journal} {Soviet Physics JETP}\ }\textbf {\bibinfo
  {volume} {42}},\ \bibinfo {pages} {164} (\bibinfo {year} {1975})}\BibitemShut
  {NoStop}%
\bibitem [{\citenamefont {Alpar}\ \emph {et~al.}(1984)\citenamefont {Alpar},
  \citenamefont {Langer},\ and\ \citenamefont {Sauls}}]{alpar1984rapid}%
  \BibitemOpen
  \bibfield  {author} {\bibinfo {author} {\bibfnamefont {M.}~\bibnamefont
  {Alpar}}, \bibinfo {author} {\bibfnamefont {S.~A.}\ \bibnamefont {Langer}}, \
  and\ \bibinfo {author} {\bibfnamefont {J.}~\bibnamefont {Sauls}},\ }\href
  {http://dx.doi.org/10.1086/162232} {\bibfield  {journal} {\bibinfo  {journal}
  {The Astrophysical Journal}\ }\textbf {\bibinfo {volume} {282}},\ \bibinfo
  {pages} {533} (\bibinfo {year} {1984})}\BibitemShut {NoStop}%
\bibitem [{\citenamefont {Sj{\"o}berg}(1976)}]{SJOBERG1976511}%
  \BibitemOpen
  \bibfield  {author} {\bibinfo {author} {\bibfnamefont {O.}~\bibnamefont
  {Sj{\"o}berg}},\ }\href {\doibase
  https://doi.org/10.1016/0375-9474(76)90558-3} {\bibfield  {journal} {\bibinfo
   {journal} {Nuclear Physics A}\ }\textbf {\bibinfo {volume} {265}},\ \bibinfo
  {pages} {511 } (\bibinfo {year} {1976})}\BibitemShut {NoStop}%
\bibitem [{\citenamefont {Borumand}\ \emph {et~al.}(1996)\citenamefont
  {Borumand}, \citenamefont {Joynt},\ and\ \citenamefont
  {Klu\ifmmode~\acute{z}\else \'{z}\fi{}niak}}]{PhysRevC.54.2745}%
  \BibitemOpen
  \bibfield  {author} {\bibinfo {author} {\bibfnamefont {M.}~\bibnamefont
  {Borumand}}, \bibinfo {author} {\bibfnamefont {R.}~\bibnamefont {Joynt}}, \
  and\ \bibinfo {author} {\bibfnamefont {W.}~\bibnamefont
  {Klu\ifmmode~\acute{z}\else \'{z}\fi{}niak}},\ }\href {\doibase
  10.1103/PhysRevC.54.2745} {\bibfield  {journal} {\bibinfo  {journal} {Phys.
  Rev. C}\ }\textbf {\bibinfo {volume} {54}},\ \bibinfo {pages} {2745}
  (\bibinfo {year} {1996})}\BibitemShut {NoStop}%
\bibitem [{\citenamefont {Baldo}\ \emph {et~al.}(1992)\citenamefont {Baldo},
  \citenamefont {Cugnon}, \citenamefont {Lejeune},\ and\ \citenamefont
  {Lombardo}}]{BALDO1992349}%
  \BibitemOpen
  \bibfield  {author} {\bibinfo {author} {\bibfnamefont {M.}~\bibnamefont
  {Baldo}}, \bibinfo {author} {\bibfnamefont {J.}~\bibnamefont {Cugnon}},
  \bibinfo {author} {\bibfnamefont {A.}~\bibnamefont {Lejeune}}, \ and\
  \bibinfo {author} {\bibfnamefont {U.}~\bibnamefont {Lombardo}},\ }\href
  {\doibase https://doi.org/10.1016/0375-9474(92)90387-Y} {\bibfield  {journal}
  {\bibinfo  {journal} {Nuclear Physics A}\ }\textbf {\bibinfo {volume}
  {536}},\ \bibinfo {pages} {349 } (\bibinfo {year} {1992})}\BibitemShut
  {NoStop}%
\bibitem [{\citenamefont {Chamel}(2008)}]{Chamel}%
  \BibitemOpen
  \bibfield  {author} {\bibinfo {author} {\bibfnamefont {N.}~\bibnamefont
  {Chamel}},\ }\href {\doibase 10.1111/j.1365-2966.2008.13426.x} {\bibfield
  {journal} {\bibinfo  {journal} {Monthly Notices of the Royal Astronomical
  Society}\ }\textbf {\bibinfo {volume} {388}},\ \bibinfo {pages} {737}
  (\bibinfo {year} {2008})}\BibitemShut {NoStop}%
\bibitem [{\citenamefont {Link}(2003)}]{link2003constraining}%
  \BibitemOpen
  \bibfield  {author} {\bibinfo {author} {\bibfnamefont {B.}~\bibnamefont
  {Link}},\ }\href {\doibase 10.1103/PhysRevLett.91.101101} {\bibfield
  {journal} {\bibinfo  {journal} {Phys. Rev. Lett.}\ }\textbf {\bibinfo
  {volume} {91}},\ \bibinfo {pages} {101101} (\bibinfo {year}
  {2003})}\BibitemShut {NoStop}%
\bibitem [{\citenamefont {Babaev}(2004)}]{babaev2004andreev}%
  \BibitemOpen
  \bibfield  {author} {\bibinfo {author} {\bibfnamefont {E.}~\bibnamefont
  {Babaev}},\ }\href@noop {} {\bibfield  {journal} {\bibinfo  {journal}
  {Physical Review D}\ }\textbf {\bibinfo {volume} {70}},\ \bibinfo {pages}
  {043001} (\bibinfo {year} {2004})}\BibitemShut {NoStop}%
\bibitem [{\citenamefont {Jones}(2006)}]{jones2006type}%
  \BibitemOpen
  \bibfield  {author} {\bibinfo {author} {\bibfnamefont {P.~B.}\ \bibnamefont
  {Jones}},\ }\href {\doibase 10.1111/j.1365-2966.2006.10754.x} {\bibfield
  {journal} {\bibinfo  {journal} {Monthly Notices of the Royal Astronomical
  Society}\ }\textbf {\bibinfo {volume} {371}},\ \bibinfo {pages} {1327}
  (\bibinfo {year} {2006})}\BibitemShut {NoStop}%
\bibitem [{\citenamefont {Alford}\ and\ \citenamefont
  {Good}(2008)}]{PhysRevB.78.024510}%
  \BibitemOpen
  \bibfield  {author} {\bibinfo {author} {\bibfnamefont {M.~G.}\ \bibnamefont
  {Alford}}\ and\ \bibinfo {author} {\bibfnamefont {G.}~\bibnamefont {Good}},\
  }\href {\doibase 10.1103/PhysRevB.78.024510} {\bibfield  {journal} {\bibinfo
  {journal} {Phys. Rev. B}\ }\textbf {\bibinfo {volume} {78}},\ \bibinfo
  {pages} {024510} (\bibinfo {year} {2008})}\BibitemShut {NoStop}%
\bibitem [{\citenamefont {Babaev}(2009)}]{PhysRevLett.103.231101}%
  \BibitemOpen
  \bibfield  {author} {\bibinfo {author} {\bibfnamefont {E.}~\bibnamefont
  {Babaev}},\ }\href {\doibase 10.1103/PhysRevLett.103.231101} {\bibfield
  {journal} {\bibinfo  {journal} {Phys. Rev. Lett.}\ }\textbf {\bibinfo
  {volume} {103}},\ \bibinfo {pages} {231101} (\bibinfo {year}
  {2009})}\BibitemShut {NoStop}%
\bibitem [{\citenamefont {Leggett}(1975)}]{RevModPhys.47.331}%
  \BibitemOpen
  \bibfield  {author} {\bibinfo {author} {\bibfnamefont {A.~J.}\ \bibnamefont
  {Leggett}},\ }\href {\doibase 10.1103/RevModPhys.47.331} {\bibfield
  {journal} {\bibinfo  {journal} {Rev. Mod. Phys.}\ }\textbf {\bibinfo {volume}
  {47}},\ \bibinfo {pages} {331} (\bibinfo {year} {1975})}\BibitemShut
  {NoStop}%
\bibitem [{\citenamefont {Chung}\ \emph {et~al.}(2007)\citenamefont {Chung},
  \citenamefont {Bluhm},\ and\ \citenamefont {Kim}}]{PhysRevLett.99.197002}%
  \BibitemOpen
  \bibfield  {author} {\bibinfo {author} {\bibfnamefont {S.~B.}\ \bibnamefont
  {Chung}}, \bibinfo {author} {\bibfnamefont {H.}~\bibnamefont {Bluhm}}, \ and\
  \bibinfo {author} {\bibfnamefont {E.-A.}\ \bibnamefont {Kim}},\ }\href
  {\doibase 10.1103/PhysRevLett.99.197002} {\bibfield  {journal} {\bibinfo
  {journal} {Phys. Rev. Lett.}\ }\textbf {\bibinfo {volume} {99}},\ \bibinfo
  {pages} {197002} (\bibinfo {year} {2007})}\BibitemShut {NoStop}%
\bibitem [{\citenamefont {Chung}\ \emph {et~al.}(2009)\citenamefont {Chung},
  \citenamefont {Agterberg},\ and\ \citenamefont {Kim}}]{chung2009fractional}%
  \BibitemOpen
  \bibfield  {author} {\bibinfo {author} {\bibfnamefont {S.~B.}\ \bibnamefont
  {Chung}}, \bibinfo {author} {\bibfnamefont {D.~F.}\ \bibnamefont
  {Agterberg}}, \ and\ \bibinfo {author} {\bibfnamefont {E.-A.}\ \bibnamefont
  {Kim}},\ }\href {http://stacks.iop.org/1367-2630/11/i=8/a=085004} {\bibfield
  {journal} {\bibinfo  {journal} {New Journal of Physics}\ }\textbf {\bibinfo
  {volume} {11}},\ \bibinfo {pages} {085004} (\bibinfo {year}
  {2009})}\BibitemShut {NoStop}%
\bibitem [{\citenamefont {Garaud}\ \emph {et~al.}(2014)\citenamefont {Garaud},
  \citenamefont {Sellin}, \citenamefont {J\"aykk\"a},\ and\ \citenamefont
  {Babaev}}]{garaud2014skyrmions}%
  \BibitemOpen
  \bibfield  {author} {\bibinfo {author} {\bibfnamefont {J.}~\bibnamefont
  {Garaud}}, \bibinfo {author} {\bibfnamefont {K.~A.~H.}\ \bibnamefont
  {Sellin}}, \bibinfo {author} {\bibfnamefont {J.}~\bibnamefont {J\"aykk\"a}},
  \ and\ \bibinfo {author} {\bibfnamefont {E.}~\bibnamefont {Babaev}},\ }\href
  {\doibase 10.1103/PhysRevB.89.104508} {\bibfield  {journal} {\bibinfo
  {journal} {Phys. Rev. B}\ }\textbf {\bibinfo {volume} {89}},\ \bibinfo
  {pages} {104508} (\bibinfo {year} {2014})}\BibitemShut {NoStop}%
\bibitem [{\citenamefont {Dahl}\ \emph
  {et~al.}(2008{\natexlab{a}})\citenamefont {Dahl}, \citenamefont {Babaev},
  \citenamefont {Kragset},\ and\ \citenamefont
  {Sudb\o{}}}]{dahl2008preemptive}%
  \BibitemOpen
  \bibfield  {author} {\bibinfo {author} {\bibfnamefont {E.~K.}\ \bibnamefont
  {Dahl}}, \bibinfo {author} {\bibfnamefont {E.}~\bibnamefont {Babaev}},
  \bibinfo {author} {\bibfnamefont {S.}~\bibnamefont {Kragset}}, \ and\
  \bibinfo {author} {\bibfnamefont {A.}~\bibnamefont {Sudb\o{}}},\ }\href
  {\doibase 10.1103/PhysRevB.77.144519} {\bibfield  {journal} {\bibinfo
  {journal} {Phys. Rev. B}\ }\textbf {\bibinfo {volume} {77}},\ \bibinfo
  {pages} {144519} (\bibinfo {year} {2008}{\natexlab{a}})}\BibitemShut
  {NoStop}%
\bibitem [{\citenamefont {Dahl}\ \emph
  {et~al.}(2008{\natexlab{b}})\citenamefont {Dahl}, \citenamefont {Babaev},\
  and\ \citenamefont {Sudb\o{}}}]{PhysRevB.78.144510}%
  \BibitemOpen
  \bibfield  {author} {\bibinfo {author} {\bibfnamefont {E.~K.}\ \bibnamefont
  {Dahl}}, \bibinfo {author} {\bibfnamefont {E.}~\bibnamefont {Babaev}}, \ and\
  \bibinfo {author} {\bibfnamefont {A.}~\bibnamefont {Sudb\o{}}},\ }\href
  {\doibase 10.1103/PhysRevB.78.144510} {\bibfield  {journal} {\bibinfo
  {journal} {Phys. Rev. B}\ }\textbf {\bibinfo {volume} {78}},\ \bibinfo
  {pages} {144510} (\bibinfo {year} {2008}{\natexlab{b}})}\BibitemShut
  {NoStop}%
\bibitem [{\citenamefont {Dahl}\ \emph
  {et~al.}(2008{\natexlab{c}})\citenamefont {Dahl}, \citenamefont {Babaev},\
  and\ \citenamefont {Sudb\o{}}}]{PhysRevLett.101.255301}%
  \BibitemOpen
  \bibfield  {author} {\bibinfo {author} {\bibfnamefont {E.~K.}\ \bibnamefont
  {Dahl}}, \bibinfo {author} {\bibfnamefont {E.}~\bibnamefont {Babaev}}, \ and\
  \bibinfo {author} {\bibfnamefont {A.}~\bibnamefont {Sudb\o{}}},\ }\href
  {\doibase 10.1103/PhysRevLett.101.255301} {\bibfield  {journal} {\bibinfo
  {journal} {Phys. Rev. Lett.}\ }\textbf {\bibinfo {volume} {101}},\ \bibinfo
  {pages} {255301} (\bibinfo {year} {2008}{\natexlab{c}})}\BibitemShut
  {NoStop}%
\bibitem [{\citenamefont {Kaurov}\ \emph {et~al.}(2005)\citenamefont {Kaurov},
  \citenamefont {Kuklov},\ and\ \citenamefont {Meyerovich}}]{kaurov2005drag}%
  \BibitemOpen
  \bibfield  {author} {\bibinfo {author} {\bibfnamefont {V.~M.}\ \bibnamefont
  {Kaurov}}, \bibinfo {author} {\bibfnamefont {A.~B.}\ \bibnamefont {Kuklov}},
  \ and\ \bibinfo {author} {\bibfnamefont {A.~E.}\ \bibnamefont {Meyerovich}},\
  }\href {\doibase 10.1103/PhysRevLett.95.090403} {\bibfield  {journal}
  {\bibinfo  {journal} {Phys. Rev. Lett.}\ }\textbf {\bibinfo {volume} {95}},\
  \bibinfo {pages} {090403} (\bibinfo {year} {2005})}\BibitemShut {NoStop}%
\bibitem [{\citenamefont {Kuklov}\ \emph
  {et~al.}(2004{\natexlab{a}})\citenamefont {Kuklov}, \citenamefont
  {Prokof'ev},\ and\ \citenamefont {Svistunov}}]{kuklov2004commensurate}%
  \BibitemOpen
  \bibfield  {author} {\bibinfo {author} {\bibfnamefont {A.}~\bibnamefont
  {Kuklov}}, \bibinfo {author} {\bibfnamefont {N.}~\bibnamefont {Prokof'ev}}, \
  and\ \bibinfo {author} {\bibfnamefont {B.}~\bibnamefont {Svistunov}},\ }\href
  {\doibase 10.1103/PhysRevLett.92.050402} {\bibfield  {journal} {\bibinfo
  {journal} {Phys. Rev. Lett.}\ }\textbf {\bibinfo {volume} {92}},\ \bibinfo
  {pages} {050402} (\bibinfo {year} {2004}{\natexlab{a}})}\BibitemShut
  {NoStop}%
\bibitem [{\citenamefont {Kuklov}\ \emph
  {et~al.}(2004{\natexlab{b}})\citenamefont {Kuklov}, \citenamefont
  {Prokof'ev},\ and\ \citenamefont {Svistunov}}]{kuklov2004superfluid}%
  \BibitemOpen
  \bibfield  {author} {\bibinfo {author} {\bibfnamefont {A.}~\bibnamefont
  {Kuklov}}, \bibinfo {author} {\bibfnamefont {N.}~\bibnamefont {Prokof'ev}}, \
  and\ \bibinfo {author} {\bibfnamefont {B.}~\bibnamefont {Svistunov}},\ }\href
  {\doibase 10.1103/PhysRevLett.92.030403} {\bibfield  {journal} {\bibinfo
  {journal} {Phys. Rev. Lett.}\ }\textbf {\bibinfo {volume} {92}},\ \bibinfo
  {pages} {030403} (\bibinfo {year} {2004}{\natexlab{b}})}\BibitemShut
  {NoStop}%
\bibitem [{\citenamefont {Svistunov}\ \emph {et~al.}(2015)\citenamefont
  {Svistunov}, \citenamefont {Babaev},\ and\ \citenamefont
  {Prokof'ev}}]{svistunov2015superfluid}%
  \BibitemOpen
  \bibfield  {author} {\bibinfo {author} {\bibfnamefont {B.~V.}\ \bibnamefont
  {Svistunov}}, \bibinfo {author} {\bibfnamefont {E.~S.}\ \bibnamefont
  {Babaev}}, \ and\ \bibinfo {author} {\bibfnamefont {N.~V.}\ \bibnamefont
  {Prokof'ev}},\ }\href@noop {} {\emph {\bibinfo {title} {Superfluid states of
  matter}}}\ (\bibinfo  {publisher} {Crc Press},\ \bibinfo {year}
  {2015})\BibitemShut {NoStop}%
\bibitem [{\citenamefont {Kuklov}\ and\ \citenamefont
  {Svistunov}(2003)}]{PhysRevLett.90.100401}%
  \BibitemOpen
  \bibfield  {author} {\bibinfo {author} {\bibfnamefont {A.~B.}\ \bibnamefont
  {Kuklov}}\ and\ \bibinfo {author} {\bibfnamefont {B.~V.}\ \bibnamefont
  {Svistunov}},\ }\href {\doibase 10.1103/PhysRevLett.90.100401} {\bibfield
  {journal} {\bibinfo  {journal} {Phys. Rev. Lett.}\ }\textbf {\bibinfo
  {volume} {90}},\ \bibinfo {pages} {100401} (\bibinfo {year}
  {2003})}\BibitemShut {NoStop}%
\bibitem [{\citenamefont {Kuklov}\ \emph {et~al.}(2006)\citenamefont {Kuklov},
  \citenamefont {Prokof'ev}, \citenamefont {Svistunov},\ and\ \citenamefont
  {Troyer}}]{kuklov2006deconfined}%
  \BibitemOpen
  \bibfield  {author} {\bibinfo {author} {\bibfnamefont {A.}~\bibnamefont
  {Kuklov}}, \bibinfo {author} {\bibfnamefont {N.}~\bibnamefont {Prokof'ev}},
  \bibinfo {author} {\bibfnamefont {B.}~\bibnamefont {Svistunov}}, \ and\
  \bibinfo {author} {\bibfnamefont {M.}~\bibnamefont {Troyer}},\ }\href
  {http://dx.doi.org/10.1016/j.aop.2006.04.007} {\bibfield  {journal} {\bibinfo
   {journal} {Annals of Physics}\ }\textbf {\bibinfo {volume} {321}},\ \bibinfo
  {pages} {1602} (\bibinfo {year} {2006})}\BibitemShut {NoStop}%
\bibitem [{\citenamefont {Babaev}()}]{babaev201547phase}%
  \BibitemOpen
  \bibfield  {author} {\bibinfo {author} {\bibfnamefont {E.}~\bibnamefont
  {Babaev}},\ }\href {https://arxiv.org/abs/cond-mat/0201547} {\bibinfo
  {journal} {arXiv preprint cond-mat/0201547}\ }\BibitemShut {NoStop}%
\bibitem [{\citenamefont {Babaev}\ \emph {et~al.}(2004)\citenamefont {Babaev},
  \citenamefont {Sudb{\o}},\ and\ \citenamefont
  {Ashcroft}}]{babaev2004superconductor}%
  \BibitemOpen
\bibfield  {journal} {  }\bibfield  {author} {\bibinfo {author} {\bibfnamefont
  {E.}~\bibnamefont {Babaev}}, \bibinfo {author} {\bibfnamefont
  {A.}~\bibnamefont {Sudb{\o}}}, \ and\ \bibinfo {author} {\bibfnamefont
  {N.}~\bibnamefont {Ashcroft}},\ }\href
  {http://dx.doi.org/10.1038/nature02910} {\bibfield  {journal} {\bibinfo
  {journal} {Nature}\ }\textbf {\bibinfo {volume} {431}},\ \bibinfo {pages}
  {666} (\bibinfo {year} {2004})}\BibitemShut {NoStop}%
\bibitem [{\citenamefont {Smiseth}\ \emph {et~al.}(2005)\citenamefont
  {Smiseth}, \citenamefont {Sm{\o}rgrav}, \citenamefont {Babaev},\ and\
  \citenamefont {Sudb{\o}}}]{smiseth2005field}%
  \BibitemOpen
  \bibfield  {author} {\bibinfo {author} {\bibfnamefont {J.}~\bibnamefont
  {Smiseth}}, \bibinfo {author} {\bibfnamefont {E.}~\bibnamefont
  {Sm{\o}rgrav}}, \bibinfo {author} {\bibfnamefont {E.}~\bibnamefont {Babaev}},
  \ and\ \bibinfo {author} {\bibfnamefont {A.}~\bibnamefont {Sudb{\o}}},\
  }\href {https://doi.org/10.1103/PhysRevB.71.214509} {\bibfield  {journal}
  {\bibinfo  {journal} {Physical Review B}\ }\textbf {\bibinfo {volume} {71}},\
  \bibinfo {pages} {214509} (\bibinfo {year} {2005})}\BibitemShut {NoStop}%
\bibitem [{\citenamefont {Herland}\ \emph {et~al.}(2010)\citenamefont
  {Herland}, \citenamefont {Babaev},\ and\ \citenamefont
  {Sudb{\o}}}]{herland2010phase}%
  \BibitemOpen
  \bibfield  {author} {\bibinfo {author} {\bibfnamefont {E.~V.}\ \bibnamefont
  {Herland}}, \bibinfo {author} {\bibfnamefont {E.}~\bibnamefont {Babaev}}, \
  and\ \bibinfo {author} {\bibfnamefont {A.}~\bibnamefont {Sudb{\o}}},\ }\href
  {https://doi.org/10.1103/PhysRevB.82.134511} {\bibfield  {journal} {\bibinfo
  {journal} {Physical Review B}\ }\textbf {\bibinfo {volume} {82}},\ \bibinfo
  {pages} {134511} (\bibinfo {year} {2010})}\BibitemShut {NoStop}%
\bibitem [{\citenamefont {Senthil}\ \emph {et~al.}(2004)\citenamefont
  {Senthil}, \citenamefont {Vishwanath}, \citenamefont {Balents}, \citenamefont
  {Sachdev},\ and\ \citenamefont {Fisher}}]{senthil2004deconfined}%
  \BibitemOpen
  \bibfield  {author} {\bibinfo {author} {\bibfnamefont {T.}~\bibnamefont
  {Senthil}}, \bibinfo {author} {\bibfnamefont {A.}~\bibnamefont {Vishwanath}},
  \bibinfo {author} {\bibfnamefont {L.}~\bibnamefont {Balents}}, \bibinfo
  {author} {\bibfnamefont {S.}~\bibnamefont {Sachdev}}, \ and\ \bibinfo
  {author} {\bibfnamefont {M.~P.}\ \bibnamefont {Fisher}},\ }\href
  {https://doi.org/10.1126/science.1091806} {\bibfield  {journal} {\bibinfo
  {journal} {Science}\ }\textbf {\bibinfo {volume} {303}},\ \bibinfo {pages}
  {1490} (\bibinfo {year} {2004})}\BibitemShut {NoStop}%
\bibitem [{\citenamefont {Kuklov}\ \emph {et~al.}(2008)\citenamefont {Kuklov},
  \citenamefont {Matsumoto}, \citenamefont {Prokof'ev}, \citenamefont
  {Svistunov},\ and\ \citenamefont {Troyer}}]{kuklov2008deconfined}%
  \BibitemOpen
  \bibfield  {author} {\bibinfo {author} {\bibfnamefont {A.~B.}\ \bibnamefont
  {Kuklov}}, \bibinfo {author} {\bibfnamefont {M.}~\bibnamefont {Matsumoto}},
  \bibinfo {author} {\bibfnamefont {N.~V.}\ \bibnamefont {Prokof'ev}}, \bibinfo
  {author} {\bibfnamefont {B.~V.}\ \bibnamefont {Svistunov}}, \ and\ \bibinfo
  {author} {\bibfnamefont {M.}~\bibnamefont {Troyer}},\ }\href {\doibase
  10.1103/PhysRevLett.101.050405} {\bibfield  {journal} {\bibinfo  {journal}
  {Phys. Rev. Lett.}\ }\textbf {\bibinfo {volume} {101}},\ \bibinfo {pages}
  {050405} (\bibinfo {year} {2008})}\BibitemShut {NoStop}%
\bibitem [{\citenamefont {Herland}\ \emph {et~al.}(2013)\citenamefont
  {Herland}, \citenamefont {Bojesen}, \citenamefont {Babaev},\ and\
  \citenamefont {Sudb{\o}}}]{herland2013phase}%
  \BibitemOpen
  \bibfield  {author} {\bibinfo {author} {\bibfnamefont {E.~V.}\ \bibnamefont
  {Herland}}, \bibinfo {author} {\bibfnamefont {T.~A.}\ \bibnamefont
  {Bojesen}}, \bibinfo {author} {\bibfnamefont {E.}~\bibnamefont {Babaev}}, \
  and\ \bibinfo {author} {\bibfnamefont {A.}~\bibnamefont {Sudb{\o}}},\ }\href
  {https://link.aps.org/doi/10.1103/PhysRevB.87.134503} {\bibfield  {journal}
  {\bibinfo  {journal} {Physical Review B}\ }\textbf {\bibinfo {volume} {87}},\
  \bibinfo {pages} {134503} (\bibinfo {year} {2013})}\BibitemShut {NoStop}%
\bibitem [{\citenamefont {Chen}\ \emph {et~al.}(2013)\citenamefont {Chen},
  \citenamefont {Huang}, \citenamefont {Deng}, \citenamefont {Kuklov},
  \citenamefont {Prokof'ev},\ and\ \citenamefont
  {Svistunov}}]{chen2013deconfined}%
  \BibitemOpen
  \bibfield  {author} {\bibinfo {author} {\bibfnamefont {K.}~\bibnamefont
  {Chen}}, \bibinfo {author} {\bibfnamefont {Y.}~\bibnamefont {Huang}},
  \bibinfo {author} {\bibfnamefont {Y.}~\bibnamefont {Deng}}, \bibinfo {author}
  {\bibfnamefont {A.~B.}\ \bibnamefont {Kuklov}}, \bibinfo {author}
  {\bibfnamefont {N.~V.}\ \bibnamefont {Prokof'ev}}, \ and\ \bibinfo {author}
  {\bibfnamefont {B.~V.}\ \bibnamefont {Svistunov}},\ }\href {\doibase
  10.1103/PhysRevLett.110.185701} {\bibfield  {journal} {\bibinfo  {journal}
  {Phys. Rev. Lett.}\ }\textbf {\bibinfo {volume} {110}},\ \bibinfo {pages}
  {185701} (\bibinfo {year} {2013})}\BibitemShut {NoStop}%
\bibitem [{\citenamefont {Sellin}\ and\ \citenamefont
  {Babaev}(2016)}]{sellin2016first}%
  \BibitemOpen
  \bibfield  {author} {\bibinfo {author} {\bibfnamefont {K.~A.~H.}\
  \bibnamefont {Sellin}}\ and\ \bibinfo {author} {\bibfnamefont
  {E.}~\bibnamefont {Babaev}},\ }\href
  {https://doi.org/10.1103/PhysRevB.93.054524} {\bibfield  {journal} {\bibinfo
  {journal} {Physical Review B}\ }\textbf {\bibinfo {volume} {93}},\ \bibinfo
  {pages} {054524} (\bibinfo {year} {2016})}\BibitemShut {NoStop}%
\bibitem [{\citenamefont {Fil}\ and\ \citenamefont
  {Shevchenko}(2005)}]{fil2005nondissipative}%
  \BibitemOpen
  \bibfield  {author} {\bibinfo {author} {\bibfnamefont {D.~V.}\ \bibnamefont
  {Fil}}\ and\ \bibinfo {author} {\bibfnamefont {S.~I.}\ \bibnamefont
  {Shevchenko}},\ }\href {\doibase 10.1103/PhysRevA.72.013616} {\bibfield
  {journal} {\bibinfo  {journal} {Phys. Rev. A}\ }\textbf {\bibinfo {volume}
  {72}},\ \bibinfo {pages} {013616} (\bibinfo {year} {2005})}\BibitemShut
  {NoStop}%
\bibitem [{\citenamefont {Linder}\ and\ \citenamefont
  {Sudb{\o}}(2009)}]{linder2009calculation}%
  \BibitemOpen
  \bibfield  {author} {\bibinfo {author} {\bibfnamefont {J.}~\bibnamefont
  {Linder}}\ and\ \bibinfo {author} {\bibfnamefont {A.}~\bibnamefont
  {Sudb{\o}}},\ }\href@noop {} {\bibfield  {journal} {\bibinfo  {journal}
  {Physical Review A}\ }\textbf {\bibinfo {volume} {79}},\ \bibinfo {pages}
  {063610} (\bibinfo {year} {2009})}\BibitemShut {NoStop}%
\bibitem [{\citenamefont {Hofer}\ \emph {et~al.}(2012)\citenamefont {Hofer},
  \citenamefont {Bruder},\ and\ \citenamefont {Stojanovi\ifmmode~\acute{c}\else
  \'{c}\fi{}}}]{hofer2012superfluid}%
  \BibitemOpen
  \bibfield  {author} {\bibinfo {author} {\bibfnamefont {P.~P.}\ \bibnamefont
  {Hofer}}, \bibinfo {author} {\bibfnamefont {C.}~\bibnamefont {Bruder}}, \
  and\ \bibinfo {author} {\bibfnamefont {V.~M.}\ \bibnamefont
  {Stojanovi\ifmmode~\acute{c}\else \'{c}\fi{}}},\ }\href {\doibase
  10.1103/PhysRevA.86.033627} {\bibfield  {journal} {\bibinfo  {journal} {Phys.
  Rev. A}\ }\textbf {\bibinfo {volume} {86}},\ \bibinfo {pages} {033627}
  (\bibinfo {year} {2012})}\BibitemShut {NoStop}%
\bibitem [{\citenamefont {Nespolo}\ \emph {et~al.}(2017)\citenamefont
  {Nespolo}, \citenamefont {Astrakharchik},\ and\ \citenamefont
  {Recati}}]{10.1088/1367-2630/aa93a0}%
  \BibitemOpen
  \bibfield  {author} {\bibinfo {author} {\bibfnamefont {J.}~\bibnamefont
  {Nespolo}}, \bibinfo {author} {\bibfnamefont {G.~E.}\ \bibnamefont
  {Astrakharchik}}, \ and\ \bibinfo {author} {\bibfnamefont {A.}~\bibnamefont
  {Recati}},\ }\href {http://stacks.iop.org/1367-2630/19/i=12/a=125005}
  {\bibfield  {journal} {\bibinfo  {journal} {New Journal of Physics}\ }\textbf
  {\bibinfo {volume} {19}},\ \bibinfo {pages} {125005} (\bibinfo {year}
  {2017})}\BibitemShut {NoStop}%
\bibitem [{\citenamefont {Fisher}\ \emph {et~al.}(1973)\citenamefont {Fisher},
  \citenamefont {Barber},\ and\ \citenamefont {Jasnow}}]{fisher1973helicity}%
  \BibitemOpen
  \bibfield  {author} {\bibinfo {author} {\bibfnamefont {M.~E.}\ \bibnamefont
  {Fisher}}, \bibinfo {author} {\bibfnamefont {M.~N.}\ \bibnamefont {Barber}},
  \ and\ \bibinfo {author} {\bibfnamefont {D.}~\bibnamefont {Jasnow}},\ }\href
  {\doibase 10.1103/PhysRevA.8.1111} {\bibfield  {journal} {\bibinfo  {journal}
  {Phys. Rev. A}\ }\textbf {\bibinfo {volume} {8}},\ \bibinfo {pages} {1111}
  (\bibinfo {year} {1973})}\BibitemShut {NoStop}%
\bibitem [{Note1()}]{Note1}%
  \BibitemOpen
  \bibinfo {note} {In some cases, e.g. \cite
  {fil2005nondissipative,svistunov2015superfluid}, \protect \textup {\hbox
  {\mathsurround \z@ \protect \normalfont (\ignorespaces \ref
  {eq:Etwospecies}\unskip \@@italiccorr )}} is written on the form $f=(\rho
  _a-\rho _{ab}) \protect \boldsymbol {v}_a^2/2 +(\rho _b-\rho _{ab})\protect
  \boldsymbol {v}_b^2/2 +\rho _{ab}\protect \boldsymbol {v}_a\cdot \protect
  \boldsymbol {v}_b$, we however find it more convenient by working in the form
  of \protect \textup {\hbox {\mathsurround \z@ \protect \normalfont
  (\ignorespaces \ref {eq:Etwospecies}\unskip \@@italiccorr )}} via the
  redefinitions $\rho _{a/b}-\rho _{ab}\rightarrow \rho _{a/b}$.}\BibitemShut
  {Stop}%
\bibitem [{\citenamefont {Onsager}(1949)}]{Onsager1949}%
  \BibitemOpen
  \bibfield  {author} {\bibinfo {author} {\bibfnamefont {L.}~\bibnamefont
  {Onsager}},\ }\href {\doibase 10.1007/BF02780991} {\bibfield  {journal}
  {\bibinfo  {journal} {Il Nuovo Cimento (1943-1954)}\ }\textbf {\bibinfo
  {volume} {6}},\ \bibinfo {pages} {279} (\bibinfo {year} {1949})}\BibitemShut
  {NoStop}%
\bibitem [{\citenamefont {Feynman}(1955)}]{FEYNMAN195517}%
  \BibitemOpen
  \bibfield  {author} {\bibinfo {author} {\bibfnamefont {R.}~\bibnamefont
  {Feynman}},\ }\href {\doibase 10.1016/S0079-6417(08)60077-3} {\ \bibinfo
  {series} {Progress in Low Temperature Physics},\ \textbf {\bibinfo {volume}
  {1}},\ \bibinfo {pages} {17 } (\bibinfo {year} {1955})}\BibitemShut {NoStop}%
\bibitem [{\citenamefont {Papp}\ \emph {et~al.}(2008)\citenamefont {Papp},
  \citenamefont {Pino},\ and\ \citenamefont {Wieman}}]{papp2008tunable}%
  \BibitemOpen
  \bibfield  {author} {\bibinfo {author} {\bibfnamefont {S.~B.}\ \bibnamefont
  {Papp}}, \bibinfo {author} {\bibfnamefont {J.~M.}\ \bibnamefont {Pino}}, \
  and\ \bibinfo {author} {\bibfnamefont {C.~E.}\ \bibnamefont {Wieman}},\
  }\href {\doibase 10.1103/PhysRevLett.101.040402} {\bibfield  {journal}
  {\bibinfo  {journal} {Phys. Rev. Lett.}\ }\textbf {\bibinfo {volume} {101}},\
  \bibinfo {pages} {040402} (\bibinfo {year} {2008})}\BibitemShut {NoStop}%
\bibitem [{\citenamefont {Thalhammer}\ \emph {et~al.}(2008)\citenamefont
  {Thalhammer}, \citenamefont {Barontini}, \citenamefont {De~Sarlo},
  \citenamefont {Catani}, \citenamefont {Minardi},\ and\ \citenamefont
  {Inguscio}}]{thalhammer2008double}%
  \BibitemOpen
  \bibfield  {author} {\bibinfo {author} {\bibfnamefont {G.}~\bibnamefont
  {Thalhammer}}, \bibinfo {author} {\bibfnamefont {G.}~\bibnamefont
  {Barontini}}, \bibinfo {author} {\bibfnamefont {L.}~\bibnamefont {De~Sarlo}},
  \bibinfo {author} {\bibfnamefont {J.}~\bibnamefont {Catani}}, \bibinfo
  {author} {\bibfnamefont {F.}~\bibnamefont {Minardi}}, \ and\ \bibinfo
  {author} {\bibfnamefont {M.}~\bibnamefont {Inguscio}},\ }\href {\doibase
  10.1103/PhysRevLett.100.210402} {\bibfield  {journal} {\bibinfo  {journal}
  {Phys. Rev. Lett.}\ }\textbf {\bibinfo {volume} {100}},\ \bibinfo {pages}
  {210402} (\bibinfo {year} {2008})}\BibitemShut {NoStop}%
\bibitem [{\citenamefont {Gersch}\ and\ \citenamefont
  {Knollman}(1963)}]{gersch1963quantum}%
  \BibitemOpen
  \bibfield  {author} {\bibinfo {author} {\bibfnamefont {H.~A.}\ \bibnamefont
  {Gersch}}\ and\ \bibinfo {author} {\bibfnamefont {G.~C.}\ \bibnamefont
  {Knollman}},\ }\href {\doibase 10.1103/PhysRev.129.959} {\bibfield  {journal}
  {\bibinfo  {journal} {Phys. Rev.}\ }\textbf {\bibinfo {volume} {129}},\
  \bibinfo {pages} {959} (\bibinfo {year} {1963})}\BibitemShut {NoStop}%
\bibitem [{\citenamefont {Jaksch}\ \emph {et~al.}(1998)\citenamefont {Jaksch},
  \citenamefont {Bruder}, \citenamefont {Cirac}, \citenamefont {Gardiner},\
  and\ \citenamefont {Zoller}}]{jaksch1998cold}%
  \BibitemOpen
  \bibfield  {author} {\bibinfo {author} {\bibfnamefont {D.}~\bibnamefont
  {Jaksch}}, \bibinfo {author} {\bibfnamefont {C.}~\bibnamefont {Bruder}},
  \bibinfo {author} {\bibfnamefont {J.~I.}\ \bibnamefont {Cirac}}, \bibinfo
  {author} {\bibfnamefont {C.~W.}\ \bibnamefont {Gardiner}}, \ and\ \bibinfo
  {author} {\bibfnamefont {P.}~\bibnamefont {Zoller}},\ }\href {\doibase
  10.1103/PhysRevLett.81.3108} {\bibfield  {journal} {\bibinfo  {journal}
  {Phys. Rev. Lett.}\ }\textbf {\bibinfo {volume} {81}},\ \bibinfo {pages}
  {3108} (\bibinfo {year} {1998})}\BibitemShut {NoStop}%
\bibitem [{\citenamefont {Greiner}\ \emph {et~al.}(2002)\citenamefont
  {Greiner}, \citenamefont {Mandel}, \citenamefont {Esslinger}, \citenamefont
  {H{\"a}nsch},\ and\ \citenamefont {Bloch}}]{greiner2002quantum}%
  \BibitemOpen
  \bibfield  {author} {\bibinfo {author} {\bibfnamefont {M.}~\bibnamefont
  {Greiner}}, \bibinfo {author} {\bibfnamefont {O.}~\bibnamefont {Mandel}},
  \bibinfo {author} {\bibfnamefont {T.}~\bibnamefont {Esslinger}}, \bibinfo
  {author} {\bibfnamefont {T.~W.}\ \bibnamefont {H{\"a}nsch}}, \ and\ \bibinfo
  {author} {\bibfnamefont {I.}~\bibnamefont {Bloch}},\ }\href
  {http://dx.doi.org/10.1038/415039a} {\bibfield  {journal} {\bibinfo
  {journal} {Nature}\ }\textbf {\bibinfo {volume} {415}},\ \bibinfo {pages}
  {39} (\bibinfo {year} {2002})}\BibitemShut {NoStop}%
\bibitem [{\citenamefont {Jaksch}\ and\ \citenamefont
  {Zoller}(2005)}]{jaksch2005cold}%
  \BibitemOpen
  \bibfield  {author} {\bibinfo {author} {\bibfnamefont {D.}~\bibnamefont
  {Jaksch}}\ and\ \bibinfo {author} {\bibfnamefont {P.}~\bibnamefont
  {Zoller}},\ }\href {\doibase https://doi.org/10.1016/j.aop.2004.09.010}
  {\bibfield  {journal} {\bibinfo  {journal} {Annals of Physics}\ }\textbf
  {\bibinfo {volume} {315}},\ \bibinfo {pages} {52 } (\bibinfo {year}
  {2005})},\ \bibinfo {note} {special Issue}\BibitemShut {NoStop}%
\bibitem [{\citenamefont {Freericks}\ and\ \citenamefont
  {Monien}(1994)}]{freericks1994phase}%
  \BibitemOpen
  \bibfield  {author} {\bibinfo {author} {\bibfnamefont {J.~K.}\ \bibnamefont
  {Freericks}}\ and\ \bibinfo {author} {\bibfnamefont {H.}~\bibnamefont
  {Monien}},\ }\href {http://stacks.iop.org/0295-5075/26/i=7/a=012} {\bibfield
  {journal} {\bibinfo  {journal} {EPL (Europhysics Letters)}\ }\textbf
  {\bibinfo {volume} {26}},\ \bibinfo {pages} {545} (\bibinfo {year}
  {1994})}\BibitemShut {NoStop}%
\bibitem [{\citenamefont {Damski}\ and\ \citenamefont
  {Zakrzewski}(2006)}]{damski2006mott}%
  \BibitemOpen
  \bibfield  {author} {\bibinfo {author} {\bibfnamefont {B.}~\bibnamefont
  {Damski}}\ and\ \bibinfo {author} {\bibfnamefont {J.}~\bibnamefont
  {Zakrzewski}},\ }\href {\doibase 10.1103/PhysRevA.74.043609} {\bibfield
  {journal} {\bibinfo  {journal} {Phys. Rev. A}\ }\textbf {\bibinfo {volume}
  {74}},\ \bibinfo {pages} {043609} (\bibinfo {year} {2006})}\BibitemShut
  {NoStop}%
\bibitem [{\citenamefont {Ejima}\ \emph {et~al.}(2012)\citenamefont {Ejima},
  \citenamefont {Fehske}, \citenamefont {Gebhard}, \citenamefont
  {zu~M\"unster}, \citenamefont {Knap}, \citenamefont {Arrigoni},\ and\
  \citenamefont {von~der Linden}}]{ejima2012characterization}%
  \BibitemOpen
  \bibfield  {author} {\bibinfo {author} {\bibfnamefont {S.}~\bibnamefont
  {Ejima}}, \bibinfo {author} {\bibfnamefont {H.}~\bibnamefont {Fehske}},
  \bibinfo {author} {\bibfnamefont {F.}~\bibnamefont {Gebhard}}, \bibinfo
  {author} {\bibfnamefont {K.}~\bibnamefont {zu~M\"unster}}, \bibinfo {author}
  {\bibfnamefont {M.}~\bibnamefont {Knap}}, \bibinfo {author} {\bibfnamefont
  {E.}~\bibnamefont {Arrigoni}}, \ and\ \bibinfo {author} {\bibfnamefont
  {W.}~\bibnamefont {von~der Linden}},\ }\href {\doibase
  10.1103/PhysRevA.85.053644} {\bibfield  {journal} {\bibinfo  {journal} {Phys.
  Rev. A}\ }\textbf {\bibinfo {volume} {85}},\ \bibinfo {pages} {053644}
  (\bibinfo {year} {2012})}\BibitemShut {NoStop}%
\bibitem [{\citenamefont {Fisher}\ \emph {et~al.}(1989)\citenamefont {Fisher},
  \citenamefont {Weichman}, \citenamefont {Grinstein},\ and\ \citenamefont
  {Fisher}}]{fisher1989boson}%
  \BibitemOpen
  \bibfield  {author} {\bibinfo {author} {\bibfnamefont {M.~P.~A.}\
  \bibnamefont {Fisher}}, \bibinfo {author} {\bibfnamefont {P.~B.}\
  \bibnamefont {Weichman}}, \bibinfo {author} {\bibfnamefont {G.}~\bibnamefont
  {Grinstein}}, \ and\ \bibinfo {author} {\bibfnamefont {D.~S.}\ \bibnamefont
  {Fisher}},\ }\href {\doibase 10.1103/PhysRevB.40.546} {\bibfield  {journal}
  {\bibinfo  {journal} {Phys. Rev. B}\ }\textbf {\bibinfo {volume} {40}},\
  \bibinfo {pages} {546} (\bibinfo {year} {1989})}\BibitemShut {NoStop}%
\bibitem [{\citenamefont {Batrouni}\ and\ \citenamefont
  {Scalettar}(1992)}]{batrouni1992world}%
  \BibitemOpen
  \bibfield  {author} {\bibinfo {author} {\bibfnamefont {G.~G.}\ \bibnamefont
  {Batrouni}}\ and\ \bibinfo {author} {\bibfnamefont {R.~T.}\ \bibnamefont
  {Scalettar}},\ }\href {\doibase 10.1103/PhysRevB.46.9051} {\bibfield
  {journal} {\bibinfo  {journal} {Phys. Rev. B}\ }\textbf {\bibinfo {volume}
  {46}},\ \bibinfo {pages} {9051} (\bibinfo {year} {1992})}\BibitemShut
  {NoStop}%
\bibitem [{\citenamefont {Prokof'ev}\ \emph
  {et~al.}(1998{\natexlab{a}})\citenamefont {Prokof'ev}, \citenamefont
  {Svistunov},\ and\ \citenamefont {Tupitsyn}}]{prokof1998worm}%
  \BibitemOpen
  \bibfield  {author} {\bibinfo {author} {\bibfnamefont {N.}~\bibnamefont
  {Prokof'ev}}, \bibinfo {author} {\bibfnamefont {B.}~\bibnamefont
  {Svistunov}}, \ and\ \bibinfo {author} {\bibfnamefont {I.}~\bibnamefont
  {Tupitsyn}},\ }\href {\doibase 10.1016/S0375-9601(97)00957-2} {\bibfield
  {journal} {\bibinfo  {journal} {Physics Letters A}\ }\textbf {\bibinfo
  {volume} {238}},\ \bibinfo {pages} {253 } (\bibinfo {year}
  {1998}{\natexlab{a}})}\BibitemShut {NoStop}%
\bibitem [{\citenamefont {Prokof'ev}\ \emph
  {et~al.}(1998{\natexlab{b}})\citenamefont {Prokof'ev}, \citenamefont
  {Svistunov},\ and\ \citenamefont {Tupitsyn}}]{prokof1998exact}%
  \BibitemOpen
  \bibfield  {author} {\bibinfo {author} {\bibfnamefont {N.~V.}\ \bibnamefont
  {Prokof'ev}}, \bibinfo {author} {\bibfnamefont {B.~V.}\ \bibnamefont
  {Svistunov}}, \ and\ \bibinfo {author} {\bibfnamefont {I.~S.}\ \bibnamefont
  {Tupitsyn}},\ }\href {\doibase 10.1134/1.558661} {\bibfield  {journal}
  {\bibinfo  {journal} {Journal of Experimental and Theoretical Physics}\
  }\textbf {\bibinfo {volume} {87}},\ \bibinfo {pages} {310} (\bibinfo {year}
  {1998}{\natexlab{b}})}\BibitemShut {NoStop}%
\bibitem [{\citenamefont {Capogrosso-Sansone}\ \emph
  {et~al.}(2008)\citenamefont {Capogrosso-Sansone}, \citenamefont {S{\"o}yler},
  \citenamefont {Prokof'ev},\ and\ \citenamefont
  {Svistunov}}]{capogrosso2008monte}%
  \BibitemOpen
  \bibfield  {author} {\bibinfo {author} {\bibfnamefont {B.}~\bibnamefont
  {Capogrosso-Sansone}}, \bibinfo {author} {\bibfnamefont {{\c{S}}.~G.}\
  \bibnamefont {S{\"o}yler}}, \bibinfo {author} {\bibfnamefont
  {N.}~\bibnamefont {Prokof'ev}}, \ and\ \bibinfo {author} {\bibfnamefont
  {B.}~\bibnamefont {Svistunov}},\ }\href {\doibase 10.1103/PhysRevA.77.015602}
  {\bibfield  {journal} {\bibinfo  {journal} {Phys. Rev. A}\ }\textbf {\bibinfo
  {volume} {77}},\ \bibinfo {pages} {015602} (\bibinfo {year}
  {2008})}\BibitemShut {NoStop}%
\bibitem [{\citenamefont {Catani}\ \emph {et~al.}(2008)\citenamefont {Catani},
  \citenamefont {De~Sarlo}, \citenamefont {Barontini}, \citenamefont
  {Minardi},\ and\ \citenamefont {Inguscio}}]{catani2008degenerate}%
  \BibitemOpen
  \bibfield  {author} {\bibinfo {author} {\bibfnamefont {J.}~\bibnamefont
  {Catani}}, \bibinfo {author} {\bibfnamefont {L.}~\bibnamefont {De~Sarlo}},
  \bibinfo {author} {\bibfnamefont {G.}~\bibnamefont {Barontini}}, \bibinfo
  {author} {\bibfnamefont {F.}~\bibnamefont {Minardi}}, \ and\ \bibinfo
  {author} {\bibfnamefont {M.}~\bibnamefont {Inguscio}},\ }\href {\doibase
  10.1103/PhysRevA.77.011603} {\bibfield  {journal} {\bibinfo  {journal} {Phys.
  Rev. A}\ }\textbf {\bibinfo {volume} {77}},\ \bibinfo {pages} {011603}
  (\bibinfo {year} {2008})}\BibitemShut {NoStop}%
\bibitem [{\citenamefont {Guertler}\ \emph {et~al.}(2008)\citenamefont
  {Guertler}, \citenamefont {Troyer},\ and\ \citenamefont
  {Zhang}}]{PhysRevB.77.184505}%
  \BibitemOpen
  \bibfield  {author} {\bibinfo {author} {\bibfnamefont {S.}~\bibnamefont
  {Guertler}}, \bibinfo {author} {\bibfnamefont {M.}~\bibnamefont {Troyer}}, \
  and\ \bibinfo {author} {\bibfnamefont {F.-C.}\ \bibnamefont {Zhang}},\ }\href
  {\doibase 10.1103/PhysRevB.77.184505} {\bibfield  {journal} {\bibinfo
  {journal} {Phys. Rev. B}\ }\textbf {\bibinfo {volume} {77}},\ \bibinfo
  {pages} {184505} (\bibinfo {year} {2008})}\BibitemShut {NoStop}%
\bibitem [{\citenamefont {S{\"o}yler}\ \emph {et~al.}(2009)\citenamefont
  {S{\"o}yler}, \citenamefont {Capogrosso-Sansone}, \citenamefont {Prokof'ev},\
  and\ \citenamefont {Svistunov}}]{soyler2009sign}%
  \BibitemOpen
  \bibfield  {author} {\bibinfo {author} {\bibfnamefont {{\c{S}}.~G.}\
  \bibnamefont {S{\"o}yler}}, \bibinfo {author} {\bibfnamefont
  {B.}~\bibnamefont {Capogrosso-Sansone}}, \bibinfo {author} {\bibfnamefont
  {N.~V.}\ \bibnamefont {Prokof'ev}}, \ and\ \bibinfo {author} {\bibfnamefont
  {B.~V.}\ \bibnamefont {Svistunov}},\ }\href
  {http://stacks.iop.org/1367-2630/11/i=7/a=073036} {\bibfield  {journal}
  {\bibinfo  {journal} {New Journal of Physics}\ }\textbf {\bibinfo {volume}
  {11}},\ \bibinfo {pages} {073036} (\bibinfo {year} {2009})}\BibitemShut
  {NoStop}%
\bibitem [{\citenamefont {Lingua}\ \emph {et~al.}(2015)\citenamefont {Lingua},
  \citenamefont {Guglielmino}, \citenamefont {Penna},\ and\ \citenamefont
  {Capogrosso~Sansone}}]{lingua2015demixing}%
  \BibitemOpen
  \bibfield  {author} {\bibinfo {author} {\bibfnamefont {F.}~\bibnamefont
  {Lingua}}, \bibinfo {author} {\bibfnamefont {M.}~\bibnamefont {Guglielmino}},
  \bibinfo {author} {\bibfnamefont {V.}~\bibnamefont {Penna}}, \ and\ \bibinfo
  {author} {\bibfnamefont {B.}~\bibnamefont {Capogrosso~Sansone}},\ }\href
  {\doibase 10.1103/PhysRevA.92.053610} {\bibfield  {journal} {\bibinfo
  {journal} {Phys. Rev. A}\ }\textbf {\bibinfo {volume} {92}},\ \bibinfo
  {pages} {053610} (\bibinfo {year} {2015})}\BibitemShut {NoStop}%
\bibitem [{\citenamefont {Guglielmino}\ \emph {et~al.}(2010)\citenamefont
  {Guglielmino}, \citenamefont {Penna},\ and\ \citenamefont
  {Capogrosso-Sansone}}]{guglielmino2010mott}%
  \BibitemOpen
  \bibfield  {author} {\bibinfo {author} {\bibfnamefont {M.}~\bibnamefont
  {Guglielmino}}, \bibinfo {author} {\bibfnamefont {V.}~\bibnamefont {Penna}},
  \ and\ \bibinfo {author} {\bibfnamefont {B.}~\bibnamefont
  {Capogrosso-Sansone}},\ }\href {\doibase 10.1103/PhysRevA.82.021601}
  {\bibfield  {journal} {\bibinfo  {journal} {Phys. Rev. A}\ }\textbf {\bibinfo
  {volume} {82}},\ \bibinfo {pages} {021601} (\bibinfo {year}
  {2010})}\BibitemShut {NoStop}%
\bibitem [{\citenamefont {Feynman}(1953)}]{feynman1953atomic}%
  \BibitemOpen
  \bibfield  {author} {\bibinfo {author} {\bibfnamefont {R.~P.}\ \bibnamefont
  {Feynman}},\ }\href {\doibase 10.1103/PhysRev.91.1291} {\bibfield  {journal}
  {\bibinfo  {journal} {Phys. Rev.}\ }\textbf {\bibinfo {volume} {91}},\
  \bibinfo {pages} {1291} (\bibinfo {year} {1953})}\BibitemShut {NoStop}%
\bibitem [{\citenamefont {Feynman}\ and\ \citenamefont
  {Hibbs}(2010)}]{feynman2010quantum}%
  \BibitemOpen
  \bibfield  {author} {\bibinfo {author} {\bibfnamefont {R.}~\bibnamefont
  {Feynman}}\ and\ \bibinfo {author} {\bibfnamefont {A.}~\bibnamefont
  {Hibbs}},\ }\href@noop {} {\emph {\bibinfo {title} {Quantum Mechanics and
  Path Integrals (Emended edition)}}}\ (\bibinfo  {publisher} {Dover
  Publications Inc., New York},\ \bibinfo {year} {2010})\BibitemShut {NoStop}%
\bibitem [{\citenamefont {Ceperley}(1995)}]{ceperley1995path}%
  \BibitemOpen
  \bibfield  {author} {\bibinfo {author} {\bibfnamefont {D.~M.}\ \bibnamefont
  {Ceperley}},\ }\href {\doibase 10.1103/RevModPhys.67.279} {\bibfield
  {journal} {\bibinfo  {journal} {Rev. Mod. Phys.}\ }\textbf {\bibinfo {volume}
  {67}},\ \bibinfo {pages} {279} (\bibinfo {year} {1995})}\BibitemShut
  {NoStop}%
\bibitem [{\citenamefont {Pollock}\ and\ \citenamefont
  {Ceperley}(1987)}]{pollock1987path}%
  \BibitemOpen
  \bibfield  {author} {\bibinfo {author} {\bibfnamefont {E.~L.}\ \bibnamefont
  {Pollock}}\ and\ \bibinfo {author} {\bibfnamefont {D.~M.}\ \bibnamefont
  {Ceperley}},\ }\href {\doibase 10.1103/PhysRevB.36.8343} {\bibfield
  {journal} {\bibinfo  {journal} {Phys. Rev. B}\ }\textbf {\bibinfo {volume}
  {36}},\ \bibinfo {pages} {8343} (\bibinfo {year} {1987})}\BibitemShut
  {NoStop}%
\bibitem [{Note2()}]{Note2}%
  \BibitemOpen
  \bibinfo {note} {To avoid confusion, we point out that this winding number is
  unrelated to the previously mentioned phase winding of vortices}\BibitemShut
  {NoStop}%
\bibitem [{\citenamefont {Galteland}\ \emph {et~al.}(2015)\citenamefont
  {Galteland}, \citenamefont {Babaev},\ and\ \citenamefont
  {Sudb{\o}}}]{galteland2015thermal}%
  \BibitemOpen
  \bibfield  {author} {\bibinfo {author} {\bibfnamefont {P.~N.}\ \bibnamefont
  {Galteland}}, \bibinfo {author} {\bibfnamefont {E.}~\bibnamefont {Babaev}}, \
  and\ \bibinfo {author} {\bibfnamefont {A.}~\bibnamefont {Sudb{\o}}},\ }\href
  {http://stacks.iop.org/1367-2630/17/i=10/a=103040} {\bibfield  {journal}
  {\bibinfo  {journal} {New Journal of Physics}\ }\textbf {\bibinfo {volume}
  {17}},\ \bibinfo {pages} {103040} (\bibinfo {year} {2015})}\BibitemShut
  {NoStop}%
\bibitem [{Note3()}]{Note3}%
  \BibitemOpen
  \bibinfo {note} {A remark concerning the co- or counterdirection of
  superfluid currents is in place to avoid confusion. If the drag interaction
  $\rho _{ab}$ in \protect \textup {\hbox {\mathsurround \z@ \protect
  \normalfont (\ignorespaces \ref {eq:Etwospecies}\unskip \@@italiccorr )}} is
  positive, the system can lower its energy by having \protect \textit {
  counterdirected} superfluid currents. However, a positive $\rho _{ab}$
  corresponds to positive $\protect \boldsymbol {W}_a\cdot \protect \boldsymbol
  {W}_b$, that is {\protect \it codirected} paths. This is not a contradiction
  since the imaginary-time paths are not the real superfluid currents, or
  loosely speaking, the quantity $d \protect \boldsymbol {r}_a / d\tau \cdot
  d\protect \boldsymbol {r}_b /d\tau $ switches sign when replacing $\tau $
  with $i t$.}\BibitemShut {Stop}%
\bibitem [{\citenamefont {Metropolis}\ \emph {et~al.}(1953)\citenamefont
  {Metropolis}, \citenamefont {Rosenbluth}, \citenamefont {Rosenbluth},
  \citenamefont {Teller},\ and\ \citenamefont
  {Teller}}]{metropolis1953equation}%
  \BibitemOpen
  \bibfield  {author} {\bibinfo {author} {\bibfnamefont {N.}~\bibnamefont
  {Metropolis}}, \bibinfo {author} {\bibfnamefont {A.~W.}\ \bibnamefont
  {Rosenbluth}}, \bibinfo {author} {\bibfnamefont {M.~N.}\ \bibnamefont
  {Rosenbluth}}, \bibinfo {author} {\bibfnamefont {A.~H.}\ \bibnamefont
  {Teller}}, \ and\ \bibinfo {author} {\bibfnamefont {E.}~\bibnamefont
  {Teller}},\ }\href {\doibase 10.1063/1.1699114} {\bibfield  {journal}
  {\bibinfo  {journal} {The Journal of Chemical Physics}\ }\textbf {\bibinfo
  {volume} {21}},\ \bibinfo {pages} {1087} (\bibinfo {year}
  {1953})}\BibitemShut {NoStop}%
\bibitem [{\citenamefont {Trefzger}\ \emph {et~al.}(2011)\citenamefont
  {Trefzger}, \citenamefont {Menotti}, \citenamefont {Capogrosso-Sansone},\
  and\ \citenamefont {Lewenstein}}]{trefzger2011ultracold}%
  \BibitemOpen
  \bibfield  {author} {\bibinfo {author} {\bibfnamefont {C.}~\bibnamefont
  {Trefzger}}, \bibinfo {author} {\bibfnamefont {C.}~\bibnamefont {Menotti}},
  \bibinfo {author} {\bibfnamefont {B.}~\bibnamefont {Capogrosso-Sansone}}, \
  and\ \bibinfo {author} {\bibfnamefont {M.}~\bibnamefont {Lewenstein}},\
  }\href {http://stacks.iop.org/0953-4075/44/i=19/a=193001} {\bibfield
  {journal} {\bibinfo  {journal} {Journal of Physics B: Atomic, Molecular and
  Optical Physics}\ }\textbf {\bibinfo {volume} {44}},\ \bibinfo {pages}
  {193001} (\bibinfo {year} {2011})}\BibitemShut {NoStop}%
\bibitem [{\citenamefont {Barkema}\ and\ \citenamefont
  {Newman}(2001)}]{barkema2001monte}%
  \BibitemOpen
  \bibfield  {author} {\bibinfo {author} {\bibfnamefont {G.}~\bibnamefont
  {Barkema}}\ and\ \bibinfo {author} {\bibfnamefont {M.}~\bibnamefont
  {Newman}},\ }\href@noop {} {\emph {\bibinfo {title} {Monte Carlo methods in
  statistical physics}}}\ (\bibinfo  {publisher} {Oxford University Press},\
  \bibinfo {year} {2001})\BibitemShut {NoStop}%
\end{thebibliography}%
\end{document}